\documentclass{CVM}
\usepackage{multirow}
\usepackage{adjustbox}
\usepackage{tabularx}
\usepackage[utf8]{inputenc}
\usepackage{longtable}
\usepackage{ltcaption}
\usepackage{multicol}
\usepackage{amsmath}
\usepackage{booktabs} 
\usepackage{hyperref}
\usepackage{xltabular}
\usepackage{xcolor}
\usepackage{afterpage}
\usepackage{stfloats}
\usepackage{amssymb,enumitem}
\usepackage{rotating}
\newcommand{\tmp}[1]{#1}

\CVMsetup{
type      = {Review Article},
doi       = {s41095-0xx-xxxx-x},
title = {A Survey on Cross-Modal Interaction Between Music and Multimodal Data},
author    = {Sifei Li$^{*1,2}$, Mining Tan$^{*1,2}$, Feier Shen$^{2,1}$, Minyan Luo$^{1,2}$, Zijiao Yin$^{2}$, Fan Tang$^{3}$, Weiming Dong$^{1,2}$\cor{}, and Changsheng Xu$^{1,2}$\\

},
runauthor = {Accepted for publication in Computational Visual Media},
abstract  = {

Multimodal learning has driven innovation across various industries, particularly in the field of music. By deeply exploiting the interaction with visual media and enhancing immersion, it not only lowers the entry barriers to music but also increases the appeal of both music and visual media. This survey aims to provide a comprehensive review of multimodal tasks related to music, outlining how music contributes to multimodal learning and offering insights for researchers seeking to expand the boundaries of computational music. Unlike visual media, which are often semantically and visually intuitive, music primarily interacts with humans through auditory perception, making its data representation inherently less intuitive. Therefore, this paper first introduces common representations of music and provides an overview of music datasets. Subsequently, we categorize cross-modal interactions between music and multimodal data into: music-driven cross-modal interactions, music-oriented cross-modal interactions, and bidirectional music cross-modal interactions. For each category, we systematically trace the development of relevant sub-tasks, analyze existing limitations, and discuss emerging trends.
Furthermore, we provide a comprehensive summary of the datasets and evaluation metrics used in multimodal tasks related to music, offering benchmark references for future research. Finally, we discuss the current challenges in cross-modal interactions involving music and propose potential directions for future research.
},
keywords  = {AI music, Multimodal learning, Music generation, Music understanding},
copyright = {The Author(s)},
}
\begin{document}
\address{MAIS, Institute of Automation, Chinese Academy of
Sciences, Beijing, 100190, China. E-mail: S. Li$^*$, lisifei2022@ia.ac.cn; M. Tan$^*$, tanmining2024@ia.ac.cn; M. Luo, luominyan2025@ia.ac.cn; W. Dong, weiming.dong@ia.ac.cn; C. Xu, csxu@nlpr.ia.ac.cn. $^*$Equal contribution.}

\address{School of Artificial Intelligence, University of Chinese
Academy of Sciences, Beijing, 100190, China. E-mail: F. Shen, shenfeier21@mails.ucas.ac.cn; Z. Jiao, yinzijiao22@mails.ucas.ac.cn.}

\address{Institute of Computing Technology, Chinese Academy of Sciences, Beijing, 100190, China. E-mail: tfan.108@gmail.com.}

\maketitle
\section{Introduction}

As an auditory art form, music not only provides pure auditory pleasure but also unlocks innovative experiences through interactions with other modalities of data. In the fields of film and animation, music enhances visual content to create emotionally impactful scenes. In songs, music combines with language, where vocals and melodies work together to convey emotions and evoke resonance. In dance performances, music guides movement, allowing dancers to convey deep emotional narratives through physical expression. This multimodal fusion expands the expressive power of music and makes it more accessible to a wider audience, supporting its spread and popularity. Within a multimodal interaction framework, music serves as a bridge between different modalities, creating immersive and emotionally engaging experiences. It plays an indispensable role in cultural expression and artistic innovation. Furthermore, this multimodal synergy opens new avenues for artificial intelligence applications in music, driving transformations in music composition, performance, and appreciation, and paving the way for novel forms of musical creativity and interaction.

Artificial intelligence (AI)-driven music and multimodal interaction is an emerging interdisciplinary field focused on integrating music with other data modalities for innovative expression. By leveraging advanced AI technologies, this interaction enables the analysis, generation, and synchronization of music with other modalities, creating richer user experiences. AI-based music and multimodal interaction can be categorized into three types:
\begin{enumerate}
\item \emph{Music-driven cross-modal interaction:}
Music serves as input to generate outputs in other modalities (e.g., visuals, text). This type of interaction relies on models with a deep understanding of musical content to generate multimodal output that helps interpret or complement music, thereby enriching its expression and enhancing user engagement.
\item  \emph{Music-oriented cross-modal interaction:}
Non-musical inputs are used to retrieve or generate music, enabling personalized and context-aware composition.
\item \emph{Bidirectional music cross-modal interaction:}
Music functions as both input and output, supporting editing, understanding, and generation across modalities, or even a more adaptive music agent. This enables deep integration and fusion between music and multimodal data, fostering new possibilities for AI-driven music composition, performance, and interactive experience.
\end{enumerate}

Music exhibits a unique level of complexity, presenting significant challenges for AI in music and multimodal data interaction. Unlike visual data, which contains intuitive spatial and color features that are easier to quantify, music involves intricate interactions of melody, rhythm, and harmony with nearly infinite variations. Its emotional and artistic intent is abstract, making it difficult to define with numerical values or text. Unlike text, which has a fixed semantic structure and grammar, music communicates through continuous sequences, where meaning arises from the whole rather than individual notes, often conveying ambiguous emotions. Music is not only a multi-track signal but also typically requires a longer duration to form a complete composition. 
\tmp{Furthermore, compared to speech,  music demands higher sampling rates and spans longer durations, resulting in significantly increased data sequences. Moreover, music relies on long-term structures (e.g., phrases) rather than short-term linguistic units, which necessitates maintaining long-term structural coherence and managing the complex interplay of polyphonic harmonies. Meanwhile, the subjective nature of musical interpretation makes aligning it with other modalities particularly challenging.}

In the early stages of AI applications in music, symbolic music was more extensively studied than music in audio form~\cite{midinet2017,dong2018musegan}. However, its reliance on music theory limited access to  researchers without a musical background. Early applications were also constrained by computational limitations, focusing primarily on single-modal music tasks and neglecting cross-modal interactions. Recent advances in computational power have enabled greater exploration of audio music and multimodal interaction, though challenges remain. Music can generally be categorized into instrumental music and songs. In instrumental music, the absence of vocal tracks simplifies modeling to some extent. However, high-quality multimodal instrumental music data remains relatively scarce on the internet, imposing additional requirements on data collection. Songs  build upon instrumental music by incorporating vocal tracks. Since songs contain linguistic information, modeling songs requires additional attention to speech-related features. Modeling songs as a whole is difficult, so many studies decompose the problem by analyzing vocals and instrumentals separately. However, developing a unified framework for both instrumental and vocal components is efficient and matches user expectations, making it a significant area of research with notable breakthroughs. Despite the availability of multimodal song data online, strict copyright restrictions hinder data construction and open sharing, posing further challenges for this field.

Our survey explores the potential and challenges of cross-modal interactions between music and multimodal data, outlining the progression from music-driven to music-oriented and bidirectional cross-modal interactions. While previous studies have primarily focused on AI applications in music, particularly AI music generation methods~\cite{dash2023ai, ji2023survey, wang2024review, ji2020comprehensive, li2025survey}, recent works include reviews of foundation models in music~\cite{ma2024foundation}, their impact on music understanding~\cite{li2024survey}, and advances in video-to-music generation~\cite{ji2025comprehensive, wang2025vision}. In contrast to prior research, this paper adopts a unique perspective by exploring the development of AI in music through the lens of cross-modal interaction between music and multimodal data. We believe that multimodal music serves as a crucial pathway toward broader accessibility to music in the future and as an important means of facilitating cultural dissemination. Currently, this field is experiencing a phase of vigorous growth and holds significant research value. This review aims to clarify the research trajectory of music-multimodal data interactions, highlight key challenges and provide insights into future trends.

\begin{sidewaysfigure*}[!htbp]
    \centering
    \includegraphics[width=1\textwidth]{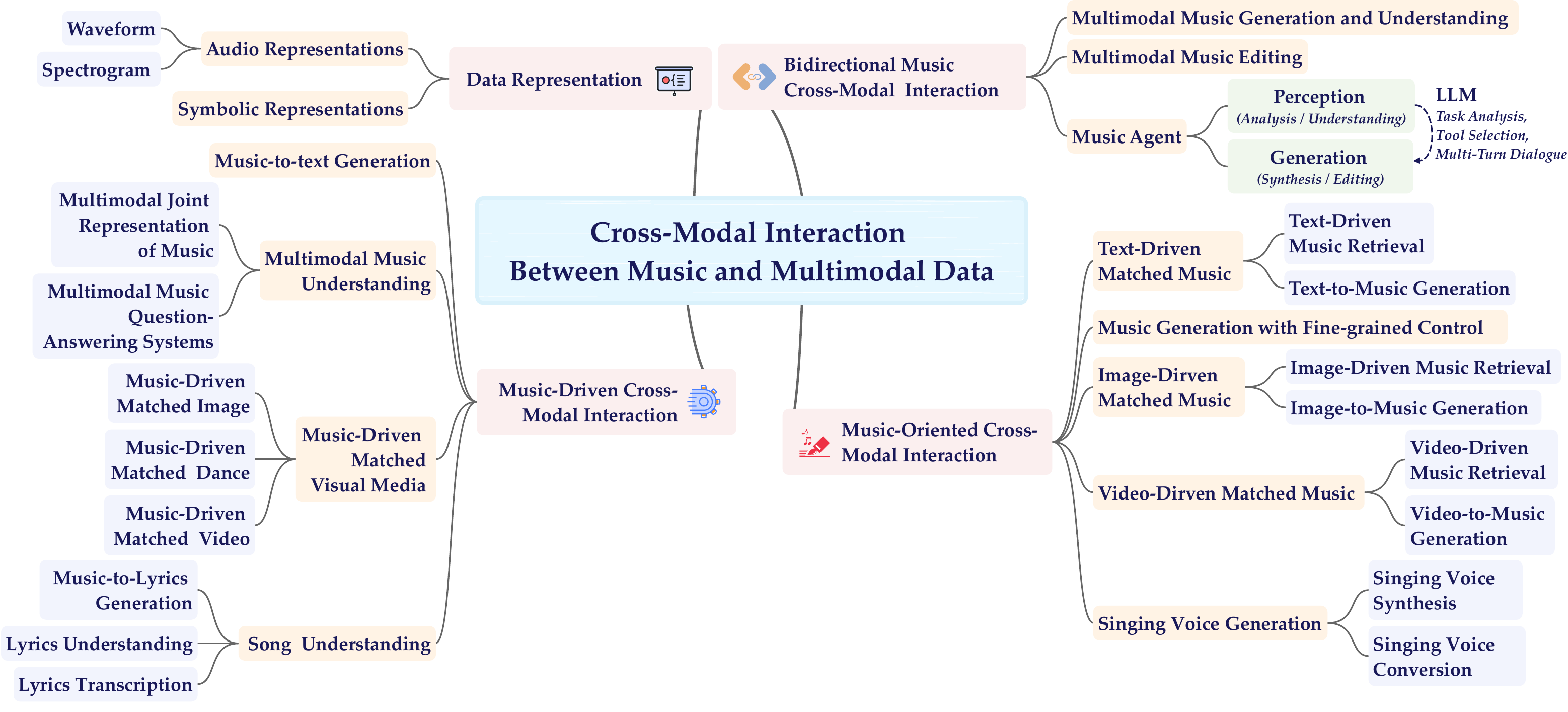}
    \caption{Overview of cross-modal interaction between music and multimodal data.}
    \label{fig:overview}
\end{sidewaysfigure*}

The structure of this review is outlined in Figure~\ref{fig:overview}. Sec.~\ref{sec:data-representation-datasets} introduces music data representations, including waveforms, spectrograms, and symbolic formats, along with related datasets. Sec.~\ref{sec:music-driven} discusses music-driven cross-modal interaction tasks and methods, while Sec.~\ref{sec:music-oriented} focuses on music-oriented cross-modal interaction tasks and methods. Sec.~\ref{sec:bidirectional-music} explores bidirectional cross-modal interaction tasks and methods. Sec.~\ref{sec:dataset-evaluation} summarizes multimodal music datasets, evaluation metrics, and their relevance to specific tasks. Finally, Sec.~\ref{sec:discussion-summary} reviews the current state of cross-modal interactions, highlights existing challenges, and suggests future research directions.

\section{Data Representation and Datasets}
\label{sec:data-representation-datasets}

Musical data representation bridges human music perception with computational models and plays a critical role in how effectively music can be processed, generated, and understood by machines. Humans perceive music through complex interactions of perception and cognition, forming features such as pitch, rhythm, dynamics, and timbre~\cite{siedenburg2019timbre, lyon2017human}. Table \ref{tab:datasets} lists several datasets commonly used to train models for music-related tasks.

\begin{table*}[p!]
\centering 
\caption{Overview of single-modal music datasets.}
\small
\begin{tabular}{>{\raggedright}p{0.16\textwidth}>{\raggedright}p{0.15\textwidth}>{\raggedright}p{0.55\textwidth}l} 

\hline
\textbf{Dataset} & \textbf{Representation} & \textbf{Description} & \textbf{Open} \\
\hline
MusicNet~\cite{thickstun2017learning} & audio & 330 classical music recordings with 1M+ annotated note-level labels, including timing, instrument, and rhythmic position of each note. & \href{https://zenodo.org/records/5120004}{Yes} \\
\hline
Million Song Dataset~\cite{bertin2011million} & audio features & Metadata and audio features of 1M songs provided by Echo Nest. & \href{http://millionsongdataset.com/}{Yes} \\ 
\hline
POP909~\cite{pop909-ismir2020} & MIDI &  909 piano arrangements of popular songs with separate melody, sub-melody and accompaniment tracks, plus tempo, rhythm, key and chord annotations. & \href{https://github.com/music-x-lab/POP909-Dataset}{Yes} \\
\hline
Free Music Archive~\cite{defferrard2016fma} & audio & 106,574 music tracks organized in a hierarchical taxonomy of 161 genres, featuring full-length high-quality audio, pre-computed features, track- and user-level metadata, tags, and free-form text. & \href{https://freemusicarchive.org/}{Yes} \\ 
\hline
GTZAN~\cite{tzanetakis_essl_cook_2001} & audio & 1,000 audio clips (each 30 seconds long), evenly distributed across 10 music genres with 100 samples per genre. & \href{http://opihi.cs.uvic.ca/sound/genres.tar.gz}{Yes} \\
\hline
Lakh MIDI Dataset~\cite{raffel2016learning} & MIDI & 176,581 MIDI files,  45,129 of which are matched and aligned with the Million Song Dataset. & \href{https://colinraffel.com/projects/lmd/}{Yes} \\
\hline
MAESTRO~\cite{hawthorne2018enabling} & MIDI, audio & Approximately 200 hours of paired audio and MIDI recordings from ten years of International Piano-e-Competition. & \href{https://magenta.withgoogle.com/datasets/maestro}{Yes} \\
\hline
ASAP~\cite{asap-dataset} & MIDI, MusicXML & 222 digital musical scores aligned with 1068 performances (more than 92 hours) of Western classical piano music. & \href{https://github.com/fosfrancesco/asap-dataset}{Yes} \\
\hline
Opencpop~\cite{wang2022opencpop} & MIDI, audio & 100 unique Mandarin songs with phonetic annotations, including utterance, note, phoneme boundaries, and pitch types, totaling 3,756 utterances lasting about 5.2 hours. & \href{https://wenet-e2e.github.io/opencpop/download/}{Yes} \\
\hline
NUS-48E Corpus~\cite{duan2013nus} & audio & 48 English pop songs by 12 singers with phoneme annotations. & \href{https://drive.google.com/drive/folders/12pP9uUl0HTVANU3IPLnumTJiRjPtVUMx}{Yes} \\
\hline
IrishMAN~\cite{DBLP:conf/hcmir/WuLY023} & ABC, MIDI, MusicXML & 216,284 Irish tunes with automatic annotations using control codes derived from ABC symbols. & \href{https://huggingface.co/datasets/sander-wood/irishman}{Yes} \\
\hline
Nottingham Music Dataset & ABC notation & 1,200 British and American folk songs. & \href{http://abc.sourceforge.net/NMD/}{Yes} \\
\hline
Henrik Norbeck's ABC Tunes & ABC notation & 3000+ Irish and Swedish folk songs. & \href{https://www.norbeck.nu/abc/}{Yes} \\
\hline
ADL piano dataset~\cite{ferreira_aiide_2020} & MIDI & 11,086 piano pieces extracted from the Lakh MIDI dataset and publicly available sources on the internet. & \href{https://github.com/lucasnfe/adl-piano-midi}{Yes} \\
\hline
Symphony~\cite{liu2022symphony} & MIDI & 46,359 files with multiple instruments and tracks, mainly orchestral. & \href{https://drive.google.com/file/d/1j9Pvtzaq8k_QIPs8e2ikvCR-BusPluTb/view?usp=sharing}{Yes} \\
\hline
Symbolic Symphony Set & MusicXML & High-quality collection of four complete symphonies (16 movements), containing 285,387 notes with detailed annotations including note events, temporal attributes, musical form, and orchestral texture. & \href{https://github.com/iis-mctl/mctl-symphony-dataset}{Yes} \\
\hline
NES-MDB~\cite{donahue2018nesmdb} & MIDI & 5,278 NES game background music tracks, featuring four programmable 8-bit register-controlled instrument voices. & \href{http://github.com/chrisdonahue/nesmdb?tab=readme-ov-file#download-links}{Yes} \\
\hline
GiantMIDI-Piano~\cite{kong2022giantmidi} & MIDI, audio & 10,855 transcribed classical piano works (1,237 hours). & \href{https://github.com/bytedance/GiantMIDI-Piano}{Yes} \\
\hline
MetaMIDI Dataset~\cite{ens2021building} & MIDI, audio & 436,631 MIDI files with metadata, matched with 10M+ Spotify music tracks. & \href{https://github.com/jeffreyjohnens/MetaMIDIDataset}{Yes} \\
\hline
MTG-Jamendo~\cite{bogdanov2019mtg} & audio & 55,000+ full audio tracks with 195 tags for genre, instrument, and mood/theme categories. & \href{https://huggingface.co/datasets/rkstgr/mtg-jamendo}{Yes} \\
\hline
Music4ALL~\cite{9145170} & audio & 109,269 tracks (911 hours) with metadata, tags, genre information, lyrics, etc. & \href{https://sites.google.com/view/contact4music4all}{Yes} \\
\hline
AudioSet & URL of audio & A large-scale collection of human-labeled 10-second sound clips drawn from YouTube videos, including 1,011,305 music clips, 117,343 musical instrument clips, etc. & \href{https://research.google.com/audioset/}{Yes} \\
\hline
AcousticBrainz & audio features & 2,524,739 audio feature files with MusicBrainz metadata. & \href{https://acousticbrainz.org/}{Yes} \\
\hline
Disco-10M~\cite{lanzendorfer2023disco} & features, URL of audio & 15296232 audio feature files covering various genres and artists. & No \\
\hline
Chordified JSB Chorales Dataset~\cite{wu2023chord} & MusicXML & 366 JSB chorales preprocessed using automated `chordification' and labelling techniques. & \href{https://github.com/sander-wood/deepchoir}{Yes} \\
\hline
GTTM Database & MusicXML & 300 classical music melodies with hierarchical music structure analysis. & \href{https://gttm.jp/gttm/database/}{Yes} \\
\hline
EMOPIA~\cite{EMOPIA} & audio, MIDI & 1,087 music clips from 387 songs and clip-level emotion labels annotated by four dedicated annotators. & \href{https://zenodo.org/record/5090631#.YPPo-JMzZz8}{Yes} \\
\hline
PopCS~\cite{liu2022diffsinger} & audio & 117 Chinese pop songs (total 5.89 hours) with lyrics. & \href{https://github.com/MoonInTheRiver/DiffSinger/blob/master/resources/apply_form.md}{Yes} \\
\hline
OpenSinger~\cite{huang2021multi} & audio & 50 hours of Mandarin singing voice recorded by 66 singers. & \href{https://drive.google.com/file/d/1EofoZxvalgMjZqzUEuEdleHIZ6SHtNuK/view?usp=sharing}{Yes} \\
\hline
M4Singer~\cite{zhang2022m4singer} & audio, MIDI & 29.77 hours of Mandarin singing voice recorded by 20 professional singers, covering all four singing styles, i.e., soprano, alto, tenor, and bass. & \href{https://m4singer.github.io/}{Yes} \\
\hline
\end{tabular}
\label{tab:datasets} 
\end{table*}

\subsection{Audio Representations}
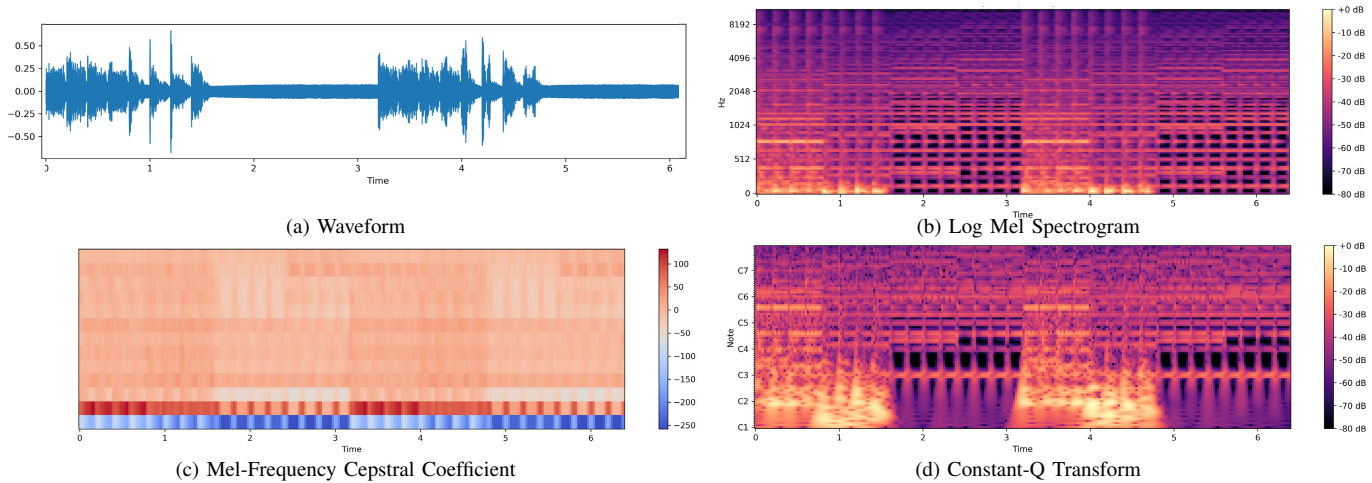
\begin{figure*}[!t]
\centering
\begin{tikzpicture}
    \node[anchor=south west,inner sep=0] (image) at (0,0)
   {\includegraphics[width=1.0\linewidth]{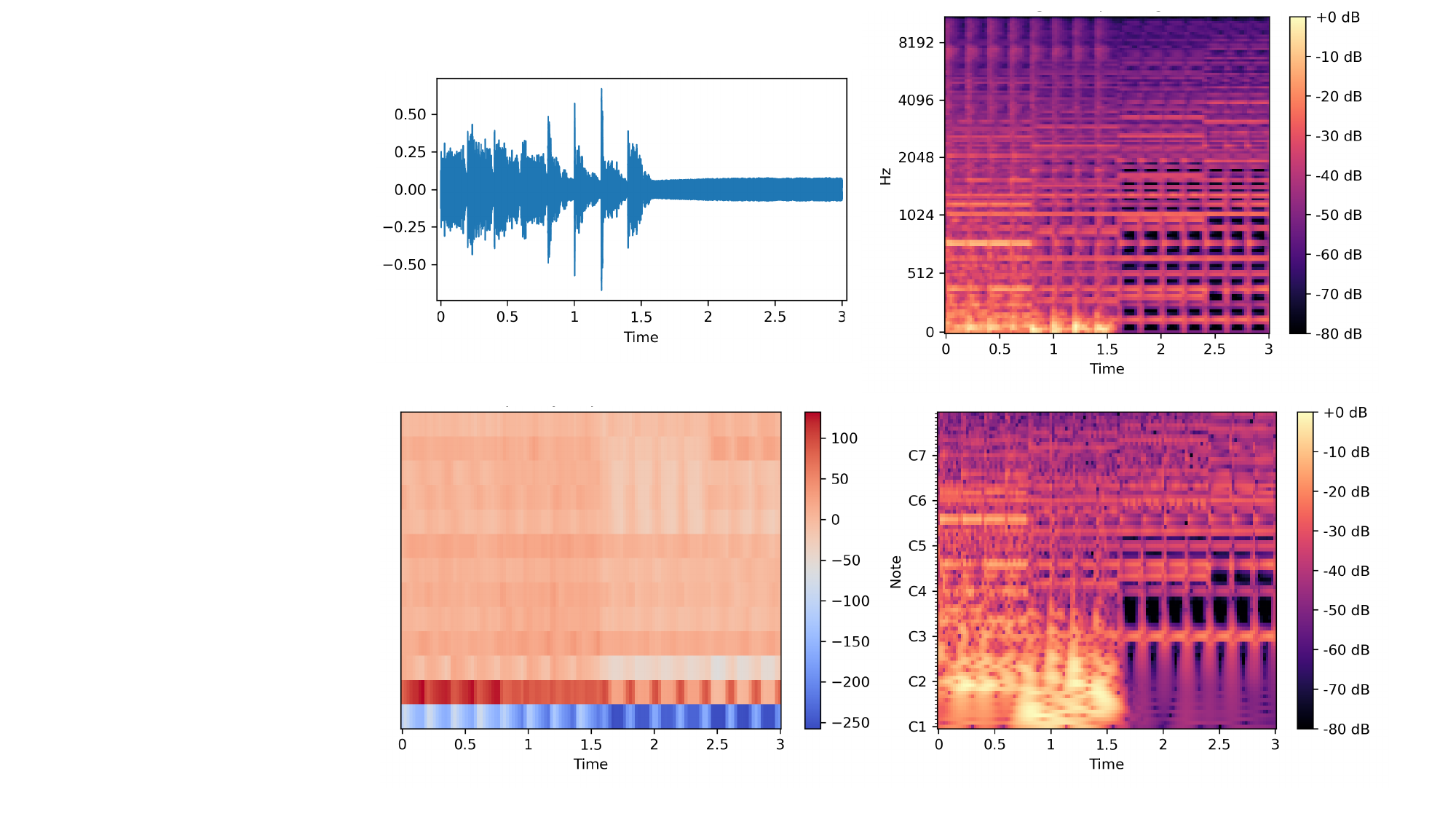}};
   \begin{scope}[x={(image.south east)},y={(image.north west)}] 
   \foreach \mylabel [count=\x from 0] in {(a) Waveform, (b) Log Mel Spectrogram} \draw (0.25 + \x * 0.52, 0.5) node {\footnotesize\mylabel};
   \foreach \mylabel [count=\x from 0] in {(c) MFCCs, (d) CQT} \draw (0.25 + \x * 0.5, -0.03) node {\footnotesize\mylabel};
    
    \end{scope}
    \end{tikzpicture}
   \caption{Examples of audio music representations}
\label{fig:spectrum}
\end{figure*}

In music information processing, audio representation is a continuous form of expression that focuses on capturing detailed acoustic features, such as timbre. Examples of audio representations are shown in Figure \ref{fig:spectrum}. 

\begin{figure*}[!t]
\centering
\begin{tikzpicture}
    \node[anchor=south west,inner sep=0] (image) at (0,0)
   {\includegraphics[width=0.8\linewidth]{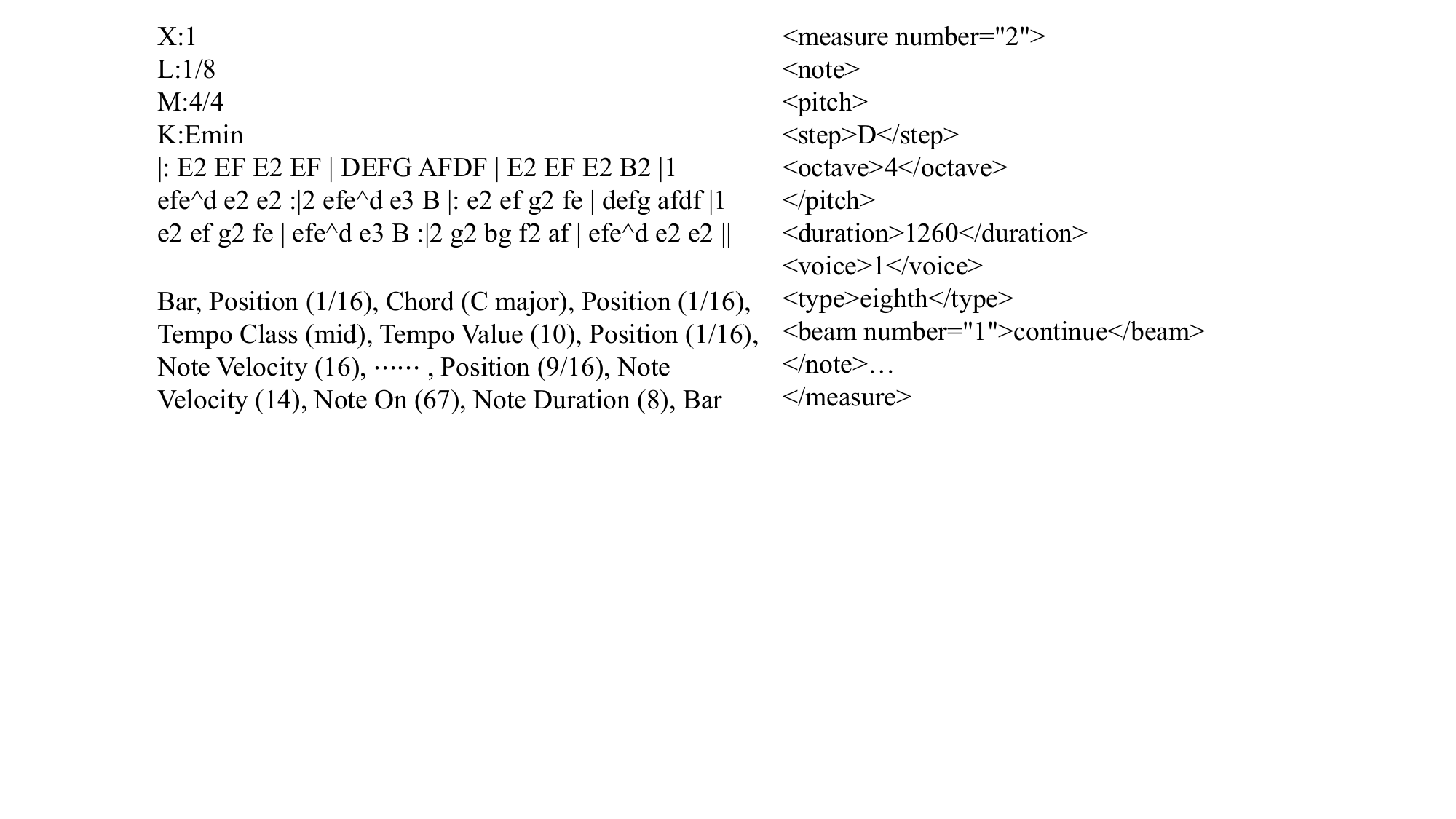}};
   \begin{scope}[x={(image.south east)},y={(image.north west)}] 

   \foreach \mylabel [count=\x from 0] in {(a) ABC notation} \draw (0.25 + \x * 0.5, 0.38) node {\footnotesize\mylabel};
   \foreach \mylabel [count=\x from 0] in {(b) REMI, (c) MusicXML} \draw (0.25 + \x * 0.55, -0.05) node {\footnotesize\mylabel};
    
    \end{scope}
    \end{tikzpicture}
   \caption{Examples of symbolic music representations}

\label{fig:sim}
\end{figure*}

The \emph{waveform} is the most direct representation of an audio signal, depicting its variation over time. The horizontal axis represents time, while the vertical axis indicates amplitude. Since waveforms precisely capture the fine details of sound signals, they serve as a fundamental representation widely used in audio processing and analysis. Many deep learning models~\cite{mehri2017samplernn, engel2017neural, goel2022s} use raw waveforms as input.

The \emph{log-mel spectrogram} is an audio representation that mimics human hearing. It starts with the short-time Fourier transform (STFT), which splits audio into overlapping frames, applies a window function, and uses FFT for time-frequency analysis. The STFT magnitude is then filtered with mel band-pass filters, designed to reflect how humans perceive sound—linear at low frequencies and logarithmic at high ones. Finally, a logarithmic transformation converts the result to a decibel scale. Although the log-mel spectrogram lowers the STFT’s spectral resolution and may lose some detail, it is widely used in audio-related tasks~\cite{hawthorne2022multi, liu2022diffsinger, gong2022ssast} due to its computational efficiency and perceptual relevance.

Using \emph{mel-frequency cepstral coefficients (MFCCs)} is a popular feature extraction and compression method in audio processing. By applying the discrete cosine transform (DCT) to the log-mel spectrum, MFCCs reduce dimensionality and inter-band correlation, retaining the most significant cepstral coefficients. The first 12-13 coefficients typically capture the key spectral information. MFCCs are commonly used in speech processing~\cite{hsu2021hubert, wang2022fusion}, and can also be applied in music classification~\cite{Yoo2024EmotionRA} and music information retrieval~\cite{yinghao2023effectiveness}.

The \emph{constant-Q transform (CQT)} refers to a filter bank in which center frequencies are exponentially spaced, the filter bandwidths vary, but the ratio of center frequency to bandwidth remains constant (Q). Unlike the Fourier transform, which uses a linear frequency axis, the CQT employs a logarithmic frequency scale that naturally follows  the musical scale. It applies longer time windows at lower frequencies for better frequency resolution and shorter windows at higher frequencies for improved time resolution, optimizing musical note frequency extraction and analysis. Although the CQT lags behind the mel spectrogram in terms of perception and reconstruction quality due to its lower frequency scaling, it has proven useful for tasks such as music representation learning~\cite{li2024mert} and musical style transfer~\cite{huang2018timbretron, demerle2024combining}.

\subsection{Symbolic Representations}
Symbolic representation encodes the logical structure of music using discrete symbols like notes, chords, and rhythms, as shown in Figure \ref{fig:sim}.

\emph{MIDI (musical instrument digital interface)} is a widely used protocol in electronic music to transmit performance data between instruments, software, and devices. Instead of encoding traditional musical notation, MIDI captures performance details such as pitch, duration, velocity, and dynamics, making it ideal for simulating human performances, though less effective at conveying musical structure. MIDI represents music as a sequence of events, including \emph{note-on, note-off, time shift, and velocity}. Its ability to capture subtle performance nuances makes it valuable for applications like music generation~\cite{lu2024musecoco}, transcription~\cite{benetos2018automatic}, and style transfer~\cite{brunner2018midi}.

\emph{REMI (revamped midi-derived events)}~\cite{huang2020pop} is an enhanced representation of MIDI data designed to address the limitations of traditional MIDI in expressing musical rhythm and structure. It introduces the \emph{note duration} event to replace the traditional \emph{note off} event, directly representing the duration of notes and enhancing the capability of rhythm modeling. Additionally, REMI employs a combination of \emph{bar} and \emph{position} tokens to replace the \emph{time shift} event, where \emph{bar} indicates the beginning of a bar and \emph{position} specifies certain positions in a bar. This design establishes a clear metric grid, facilitating more accurate modeling of musical structure. To further enhance the expressiveness of musical rhythm, REMI adds \emph{tempo} events placed before  corresponding \emph{position} events. This approach is widely used in music generation tasks~\cite{lu2024musecoco}.

\emph{MusicXML} is a notation format based on extensible markup language (XML), designed to represent various symbols used in Western music---like rests, slurs, beams, barlines, key and time signatures, articulations, and ornaments---that are often missing in MIDI. It is mainly used to store lead sheets, which include the basic parts of a song, including melodies, lyrics, chords, and repetition marks. The chords can include both literal chord symbols (e.g.\ Gmaj7) and functional notation relative to the key (e.g.\ VI7). MusicXML uses a hierarchical structure with nested tags to organize musical data. For instance, a \texttt{<measure>} contains \texttt{<note>} tags, and each \texttt{<note>} includes \texttt{<pitch>} and \texttt{<duration>}. This format helps computers understand how musical elements are related. However, its detailed structure and XML's verbosity make it challenging to use for training music generation models, especially since it lacks direct audio data and is complex to encode and decode.

The \emph{piano roll} is a common symbolic way to represent music, originating from player pianos that used perforated paper rolls to play songs automatically. In modern digital applications, the piano roll appears as a two-dimensional grid: time runs along the horizontal axis, and pitch runs along the vertical axis. Notes are shown as rectangles whose lengths indicate duration, making this format an intuitive tool for music analysis and generation. Mathematically, the piano roll is structured as an $h \times w$ real-valued matrix, where $h$ denotes the pitch range, including special entries for rests and note continuations, and $w$ represents the number of time steps, set by the beat resolution. Traditional piano rolls use binary values to show whether a note is played, but modern versions can include extra details like velocity, articulation, and dynamics~\cite{zhao2019emotional}. This image-like format has proven useful for tasks such as music generation~\cite{li2024diff, min2023polyffusion} and representation learning~\cite{wang2020pianotree}.

\emph{ABC notation} is a concise, computer-friendly way of representing music using the ASCII character set. It uses the letters A–G and a–g for notes, while ‘z’ represents a rest. The notation is structured into two main components: a header and a body. The header contains metadata like title, time signature, key signatures, and default note length. In the body, each note is represented as a token encoding pitch, duration, and rhythmic position, separated by bar lines. ABC notation's compatibility with natural language representations has made it a valuable resource for text-to-symbolic music models, facilitating the generation of music~\cite{wang2025notagen, chatmusician2024, qu2024mupt}.

\emph{Note graphs} provide a graph-based model of musical scores that capture the complex relationships in music using structured graph elements. In the method proposed by Jeong et al.~\cite{jeong2019graph}, each note is a node, and edges represent musical connections such as sequence, rest, onset, sustain, voice, and slur. These edges are directed both forward and backward, with self-loops for each note, resulting in twelve edge types. This structure offers a detailed way to represent musical scores. Later works have built on this idea. Karystinaios et al.~\cite{karystinaios2022cadence} included both notes and rests as nodes, linking them through temporal relationships like synchronization, succession, and overlap. In a follow-up study~\cite{karystinaios2023musical}, they proposed a heterogeneous graph framework for voice separation using link prediction. Similarly, Zhang et al.~\cite{zhang2023symbolic} introduced a new graph representation for musical performances and compared it to matrix and sequence-based methods in piece-level classification tasks. These studies show how graph-based models can offer richer insights than traditional music representations.

\section{Music-Driven Cross-Modal Interaction}
\label{sec:music-driven}
\subsection{Music-to-Text Generation}
Music-to-text generation lies at the intersection of music information retrieval and natural language processing (NLP), aiming to generate rich textual representations for music. This task can be broadly categorized into three subdomains: \emph{music understanding}, \emph{song understanding}, and \emph{lyrics generation}. 
\begin{figure*}[!t]
\centering
   \includegraphics[width=0.9\linewidth]{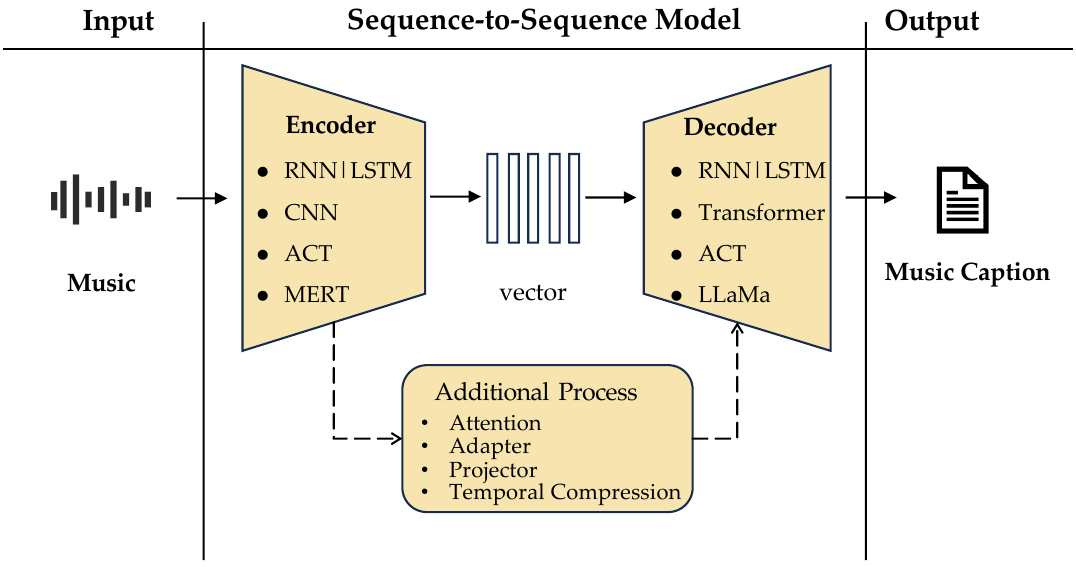}
   \caption{
   Overview of  music caption methods.}

\label{fig:music_caption}
\end{figure*}
\subsubsection{Music Understanding}

Music understanding aims to extract semantic information from music and represent it in natural language. A fundamental task in this field is music captioning, which generates coherent textual descriptions of musical segments, capturing genre, emotion, and style to enhance music retrieval and recommendation. 
\tmp{ Music captioning models typically follow an encoder-decoder framework, where the encoder extracts latent features from the raw audio, and the decoder generates the textual description based on these features (see Figure~\ref{fig:music_caption}).}

Choi et al.~\cite{choi2016towards} first explored this task, proposing a sequence-to-sequence model that encoded musical features into vectors and decoded them into descriptive text. Building on this work, Manco et al.~\cite{manco2021muscaps} proposed the MusCaps model, which employed an encoder-decoder architecture integrating convolutional neural networks (CNNs) and recurrent neural networks (RNNs), improving audio-text alignment and capturing musical nuances. Xu et al.~\cite{xu2020crnn} enhanced the model with reinforcement learning, improving caption quality for complex audio.

Despite these advances, traditional CNN-RNN architectures face limitations in temporal modeling and long-range dependency capture. To address these shortcomings, Chen et al.~\cite{chen2020audio} introduced a Transformer-based model, utilizing a CNN10 encoder to extract spectral features and a Transformer decoder to generate natural text. Mei et al.~\cite{mei2021audio} took this further with the Audio Captioning Transformer (ACT), a fully Transformer-based model using self-attention for audio processing and global context modeling. 

Recently, large language models (LLMs) have driven advances in music understanding frameworks. MusiLingo~\cite{deng2024musilingo} and MU-LLaMA~\cite{liu2024music} integrated the audio model MERT and LLaMA~\cite{touvron2023llama}, improving efficiency, accuracy, and generalization in music understanding tasks. Zhao et al.~\cite{zhao2024openmu} proposed OpenMU, a multimodal large language model for music understanding and OpenMU-Bench, a large-scale benchmark for comprehensive evaluation. These models demonstrated excellent performance in tasks like music captioning, reasoning, and question answering.

Data scarcity remains a critical challenge in this domain. Doh et al.~\cite{doh2023lp} proposed LP-MusicCaps, using an LLM to generate pseudo-music captions from extensive tagged datasets, supporting captioning model training. To improve caption diversity, Augment, Drop \& Swap~\cite{manco2024augment} optimizes the learning of music-text representations through targeted augmentation, deletion, and substitution operations, enhancing lexical variety and semantic expressiveness. Srivatsan et al.~\cite{srivatsan2023retrieval} incorporated multimodal inputs, such as video frames or musical themes, to construct context-specific prefixes, enabling captions that reflect the music’s background and emotion. Similarly, RECAP~\cite{DBLP:journals/corr/abs-2212-10901} retrieved similar music captions to construct a prompt, enhancing adaptability and precision for complex music.

Despite significant progress in music understanding, challenges such as data scarcity, coarse and simplistic music descriptions, and limited generalization in complex scenarios hinder its development. In the future, compatibility with diverse music representations, flexible question-answering with reasoning capabilities, and support for cross-cultural music are expected to become key trends.

\subsubsection{Song Understanding}

Song understanding focuses on the tasks of automatic lyrics transcription and alignment, aiming to extract lyrics from audio signals and align them precisely with the corresponding melody.

Early approaches predominantly relied on traditional audio feature analysis and handcrafted algorithms, such as hidden Markov models (HMMs)~\cite{fujihara2011lyricsynchronizer, mesaros2008automatic}, Gaussian mixture models (GMMs)~\cite{fujihara2007music}, and probabilistic models~\cite{dzhambazov2017knowknowledge}.
Although these methods succeed in constrained scenarios, manually designed features and limited model expressiveness make them unsuitable for complex audio.

The advent of deep learning revolutionized lyrics transcription and alignment, with CNNs~\cite{bittner2017deep} and RNNs~\cite{vaglio2020multilingual} demonstrating particular efficacy in handling complex musical patterns and temporal dependencies. 
Gupta et al.~\cite{gupta2020automatic} trained polyphonic acoustic models using genre-specific characteristics, improving lyrics transcription and alignment. An end-to-end model based on Wave-U-Net~\cite{stoller2019end}, utilizing connectionist temporal classification (CTC) loss, substantially enhanced alignment accuracy in multi-instrumental music. MSTRE-Net~\cite{demirel2021mstre} introduced a novel multi-streaming time-delay neural network (MTDNN) architecture, optimizing both speed and accuracy. Gao et al.~\cite{gao2022automatic} developed a multi-task learning network for lyrics transcription in polyphonic music, jointly training lyrics and chord transcription, improving transcription accuracy.

With the advent of multimodal large models, lyrics transcription and alignment progressed toward greater complexity and precision. OpenAI’s Whisper model, trained with weakly supervised learning, excels in long-form audio transcription. LyricWhiz~\cite{zhuo2023lyricwhiz} integrated Whisper with GPT-4 for zero-shot multilingual lyrics transcription, using Whisper for audio transcription and GPT-4 for text refinement. SongComposer~\cite{ding2024songcomposer} employed a two-stage approach—pre-training and supervised fine-tuning—to generate aligned lyrics-melody pairs. Addressing multilingual automatic lyrics transcription, Huang et al.~\cite{huang2024towards} proposed an end-to-end system that expanded vocabularies using existing datasets and integrated linguistic information, demonstrating the superiority of multilingual models over monolingual ones. SongTrans~\cite{wu2024songtrans} introduced a unified model for simultaneous lyrics and note transcription and alignment without preprocessing. Gu et al.~\cite{gu2024automatic} developed a multimodal framework combining audio and lip movement data to enhance lyrics and note transcription. In addition, they introduced the first multimodal singing datasets, N20EMv1 and N20EMv2.

Additionally, several studies~\cite{sterckx2017break, Ogihara2021GraphBasedRA, tadokoro2024visualization} have explored automatic lyrics interpretation, generating accessible paraphrases to deepen song comprehension. While these methods do not directly address melody, they offer fresh perspectives on musical semantics. Multimodal approaches integrating lyrics and audio can yield more accurate interpretations. For instance, Bart-fusion~\cite{zhang2022interpreting} used cross-modal attention to fuse audio and lyrics, producing more precise explanations.

Song understanding tasks currently face challenges in robust multilingual transcription and accurate alignment in complex audio. 
Future trends include unified frameworks based on multimodal pretrained models, cross-lingual alignment, and real-time transcription, to improve lyrics transcription and semantic understanding.

\subsubsection{Lyrics Generation}

Lyrics generation aims to produce rhythmically aligned and semantically coherent lyrics based on melodies or other inputs. 
This task has significant potential in music composition, entertainment, and personalized content creation, such as generating lyrics for melodies or assisting songwriters.

Early-stage approaches~\cite{barbieri2012markov, malmi2016dopelearning} primarily relied on rule-based and statistical models, utilizing only lyrics, which results in limited generation quality.

To improve alignment between lyrics and melodies, the melody-to-lyrics (M2L) approach has been widely explored. Watanabe et al.~\cite{watanabe2018melody} constructed a melody-lyrics alignment dataset and proposed an RNN-based melody-conditioned lyrics language model to generate lyrics compatible with melodies. Chen et al.~\cite{chen2020melody} developed an end-to-end system based on SeqGAN, generating lyrics from melody inputs while supporting thematic conditions to improve semantic expression. SongMASS~\cite{sheng2021songmass} addressed data scarcity and alignment constraints in lyrics-to-melody and melody-to-lyrics generation through masked sequence-to-sequence pre-training and attention-based alignment mechanisms. AI-Lyricist~\cite{ma2021ai} targeted multi-channel music and lexical constraints, producing lyrics suitable for language learning.

Subsequent studies have further advanced this field. LOAF-M2L~\cite{ou2023loaf} jointly learned wording and formatting, improving syllable compatibility between lyrics and melodies. Tian et al.~\cite{tian2023unsupervised} proposed an unsupervised hierarchical model that separated training (based on plain text) from inference (melody-guided text generation), achieving high-quality lyrics generation in unsupervised settings. UniLG~\cite{qian2023unilg} integrated text and music information through composite templates, effectively capturing the structural information of lyrics and generating content  aligned with song structures (e.g.\ chorus and verse). Zhang et al.~\cite{zhang2024controllable} introduced a note-level melody-based lyrics generation method, utilizing a Transformer architecture to capture fine-grained syllable-level correlations between melody and lyrics data, facilitating controllable and diverse lyrics generation. Calzolari et al.~\cite{chen2024scansion} proposed a lyrics generation approach based on scansion analysis, employing the mBart model to treat lyrics generation as a translation task, producing lyrics that matched both melody and tonal patterns. XAI-Lyricist~\cite{liang2024xai}, built on Transformer, provided explanations for prosodic patterns, addressing challenges in evaluating ability to be sung and alignment. Chae et al.~\cite{chae2024song} introduced a song-form-aware framework for full-song text-to-lyrics generation, enabling multi-granularity syllable count control. 
Beyond M2L lyrics generation, auxiliary lyric creation systems~\cite{zhang2020youling, zhang2022qiuniu} have also matured, assisting creators in better controlling lyric content, rhyme schemes, and formatting.

Lyrics generation tasks face several challenges, including poor cross-lingual prosody adaptation and limited ability to be sung of generated lyrics. Future works are expected to include multimodal pretrained models for lyrics generation, cross-lingual prosody modeling, and interactive user-driven creation systems, aiming to enhance the alignment between lyrics and melody.

\subsection{Multimodal Music Understanding}
Multimodal music understanding differs significantly from unimodal music understanding. Unimodal music understanding primarily focuses on music information retrieval (MIR) tasks, such as chord detection, key recognition, melody estimation, pitch detection, automatic music transcription, instrument classification, and music categorization. These tasks only take music as input and analyze specific properties of the music.
In contrast, multimodal music understanding aims to enhance the comprehension of music by integrating multiple modalities. It not only enables a deeper analysis of the intrinsic features of music but also achieves an understanding of the relationships between music and other modalities.

\subsubsection{Multimodal Representation Learning with Music}
Multimodal music representation encodes various modalities into a shared feature space, critical for understanding music and evaluating cross-modal alignment.
Among these, music-text joint representation is the most extensively studied. 
Huang et al.~\cite{huang2020large} leveraged textual tags and co-listening statistics from large-scale music video data, aligning text and music embeddings through triplet loss and cross-entropy loss. However, it still failed to understand unstructured text. Won et al.~\cite{emotion2021} explored classification, regression, and metric learning to align free-form text with music. But they found that relying solely on existing emotion labels was insufficient for modality alignment.
MuLaP~\cite{learning2021} employed weak language supervision by using masked language modeling, masked audio modeling, and audio-text matching as pretraining objectives for multimodal pretraining of audio and language data. However, its cross-attention-based modality fusion limited transfer learning applications, making it less flexible than shared embedding spaces. MuLan~\cite{Mulan2022}, inspired by CLIP, constructed a joint embedding space for music and text using large numbers of music-
text pairs. It exhibited strong representation capabilities and performed well in zero-shot downstream tasks, but the model remained closed-source.
Unlike the above approaches, which focused on audio representations, CLaMP2~\cite{wu2024clamp} separately designed a multilingual text encoder and a multimodal music encoder. Using contrastive learning, it jointly represented text in 101 languages with ABC notation and MIDI multimodal music data, demonstrating effective applications for traditional MIR tasks.

Many audio-text joint embedding models also possess some degree of music-text joint representation capability, as their datasets contain music. Favory et al.~\cite{favory2020coala, favory2021learning} employed contrastive learning to model the latent representations of audio and labels. However, their approach did not generalize to free-form natural language.
A major limitation of audio-text joint representation is the lack of paired audio-text data. CLAP~\cite{large2022} addressed this by collecting the LAION-Audio-630K dataset and applying contrastive learning to obtain a stronger audio-text joint embedding space, which efficiently handled audio of varying lengths and was widely used in downstream tasks and evaluation. Zhu et al.~\cite{Cacophony2024} constructed an even larger dataset and leveraged large language models (LLMs) to improve the quality of textual descriptions. They employed masked autoencoders (MAE) as audio encoders, achieving superior performance in downstream tasks.
Wu et al.~\cite{audio2023} revealed limitations in existing audio–text models, especially in capturing the sequential and concurrent relationships of sound events. They proposed a Transformer-based architecture to improve temporal and contextual understanding.

The above works mainly focused on the semantic alignment of audio and text. 
Some studies have focused on audio-video representation learning. CAV-MAE and MAViL~\cite{huang2024mavil} used MAE and contrastive learning to achieve audio-video joint representation. EquiAV~\cite{kimequiav} applied equivariant learning to audiovisual contrastive learning, effectively improving representation learning performance through a unique design. Diff-Feloy~\cite{luo2024diff} introduced CAVP, a model that enhanced semantic and temporal alignment in audio-video contrastive learning.

While the above studies  focused on aligning two modalities, some research has explored joint representation across all modalities. Several approaches extended existing models to make them adaptable to other modalities. For example, AudioCLIP~\cite{Audioclip2021} and Wav2CLIP~\cite{Wav2CLIP2021} integrated audio encoders into CLIP, enabling joint representation of text, images, and audio. ImageBind~\cite{girdhar2023imagebind} learned six-modality joint embeddings by leveraging image-paired data and pretrained models and using image embeddings as a bridge. VIT-LENS~\cite{vitlens2023} extended pretrained ViT models to multiple modalities and aligned their output features with anchor data features.
Another approach involved constructing multimodal datasets. Chen et al.~\cite{chen2023vast} created the VAST-27M dataset, a large multimodal video dataset, and trained a highly generalizable multimodal joint representation model, VAST.
Unlike others, CLAMP 3~\cite{wu2025clamp} paid more attention to music; it is a framework for MIR across multiple modalities and languages.

\subsubsection{Multimodal Music Question-Answering Systems}
With the emergence of LLMs, music understanding tasks have moved beyond structured text or predefined labels, enabling free-form textual outputs. Mu-LLaMa created question-answer pairs from existing audio description datasets, constructed the MusicQA dataset, and designed a music understanding adapter module to integrate music representations with LLMs, achieving efficient music question-answering and description generation.
Building on this, MusiLingo further explored music-language model alignment by constructing the MusicInstruct dataset based on MusicCaps. It used a single projection layer to align music models with language models, achieving better performance through pretraining and instruction fine-tuning. LLARK~\cite{LLark2024} integrated existing open-source datasets and employed a generative music encoder with a simple multimodal projection module, enabling music understanding, captioning, and reasoning.

To evaluate these advances, MuChoMusic~\cite{MuChoMusic2024} introduced the first benchmark dataset for assessing audio-language models' music understanding capabilities, containing 1,187 multiple-choice questions. Compared to previous works that relied on synthetic data, this benchmark offered greater reliability.
Zhao et al.~\cite{zhao2024openmu} proposed OpenMU-Bench, which expanded existing datasets to millions of training samples and introduced two additional subtasks: lyrics understanding and music tool usage. 
MuMu-LLaMA~\cite{liu2024mumu} extended multimodal musicQA to the visual domain, enabling the model to understand relationships between music, images, and video, supporting both music understanding and music generation.
Recently, several more general multimodal audio question-answering systems~\cite{Qwen-Audio2023, LTU2024, chu2024qwen2} have demonstrated reliable music understanding capabilities.

Multimodal music understanding is vital for deep music analysis, supporting the evaluation of music generation models and future systems with reasoning abilities.
Future work needs to further explore the intrinsic correspondence between multimodal data and music, enhance the flexibility of model input (e.g.\ in terms of data formats, question-answering modes), and develop chain-of-thought (CoT) data to enhance music understanding.

\subsection{Music-Driven Matched Visual Media} 
\label{sec:m2v}
Collaborative creation between music and visual content (images, videos) is a vital area in multimodal research. Next we explore how music drives  generation, alignment, and retrieval of visual content. These technologies unlock new possibilities for applications such as automatic music visualization, music video (MV) creation, and dance generation.

\subsubsection{Music-Driven Matched Image} 
Music-guided image generation has typically relied on other modalities as intermediate steps.
Specifically, music is first processed by LP-Music-Caps to generate audio captions, which are then converted into image descriptions by LLaMA2~\cite{touvron2023llama2openfoundation}, and images are generated using Stable Diffusion XL~\cite{podell2023sdxlimprovinglatentdiffusion}.

Music-driven image generation is challenging due to the stark contrast between music and images. Music is sequential, with dimensions like melody, rhythm, and harmony, while images are spatial, made up of pixels and visual elements. This gap complicates data acquisition and evaluation, and makes achieving real-world applications difficult.

\subsubsection{Music-Driven Matched Dance} 
\begin{figure*}[!t]
\centering
   \includegraphics[width=1\linewidth]{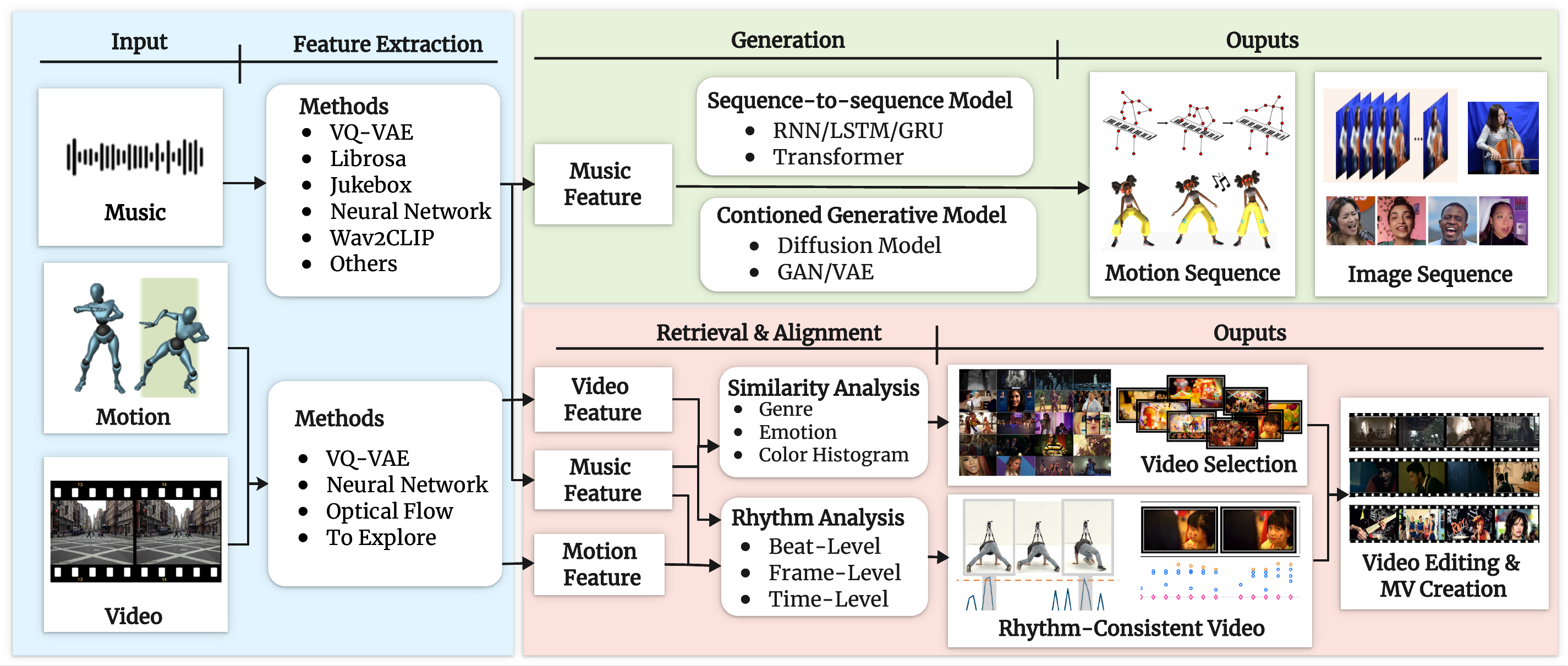}
   \caption{
   Overview of the music-driven matched \tmp{dance/video} methods.}

\label{fig:m2v}
\end{figure*}

 Dance-related research occupies a significant portion of music-driven visual media studies and is therefore discussed separately. This section is divided into two parts: music-driven dance alignment and music-driven dance generation\tmp{, as shown in Figure~\ref{fig:m2v}}.

Music-driven dance alignment refers to the adjustment of the relative timing between music and dance clips to create a rhythm-consistent dance video. Davis et al.~\cite{Davis_2018_CVPR_Workshops} introduced the concept of visual beats based on handcrafted features, and aligned them to musical beats by time-warping. AlignNet~\cite{Wang2020AlignNetAU} used a spatiotemporal attention module to extract music and video feature pyramids, and achieved dance-music alignment via frame-level correspondence estimation based on feature similarity metrics. MuDaR~\cite{Yu2022LearningMR} utilized an attention-based fully connected layer to extract motion rhythm and facilitate rhythm synchronization using both explicit and implicit methods. M2BNet~\cite{Jiang2024M2BeatsWM}, comprising MFE blocks with SGCN and TCN, effectively extracted intricate motion rhythms from videos by analyzing human skeletons and incorporating temporal information, thereby facilitating beat-level alignment.

Music-driven dance generation (also known as AI choreography) takes a music sequence as input and outputs a 3D human pose sequence, where human poses are typically represented by the rotation matrices of 24 joints and the root translation, in  SMPL format. Early works explored various methods, based on CNNs~\cite{cnns-dance1,cnns-dance2}, gated recurrent units (GRUs)~\cite{GRUs-dance}, long short-term memory (LSTM)~\cite{GrooveNet2017,TangJM18,Lstm-dance,zhuang2020music2dancedancenetmusicdrivendance},  generative adversarial networks (GANs)~\cite{dancingmusic2019,Kim_2022_CVPR}, and Transformers~\cite{AIST,Bailando,Li2021DanceFormer}. Li et al.~\cite{AIST} proposed a full-attention cross-modal transformer network that encoded seed motion and music features to generate new dance sequences autoregressively. DanceFormer~\cite{Li2021DanceFormer} employed a two-stage approach, generating key poses for each beat and then predicting  intermediate motion curves. 
\tmp{Integrating professional domain knowledge, ChoreoMaster~\cite{chen2021choreomaster} incorporated choreography rules into a graph-based framework, leveraging cross-modal embeddings to optimize motion paths for highly controllable dance synthesis.}
Bailando employed VQ-VAE to encode dance sequences \tmp{into hierarchical codebooks and designed an actor-critic GPT to generate motions by composing codebook units. To ensure rhythmic synchrony, it optimized a composite reward function: $R = \lambda_1 R_{\mathrm{beat}} + \lambda_2 R_{\mathrm{div}},$ where $R_{\mathrm{beat}}$ aligned motion kinetic peaks with musical beats and $R_{\mathrm{div}}$ promoted choreographic diversity.}
Based on Bailando, GTN-Bailando~\cite{zhuang2023gtnbailando} employed a genre token network (GTN) to identify music genres, guiding dance generation with genre consistency.
To address the limited generalization ability of GTN-Bailando, \tmp{EnchantDance~\cite{han2023enchantdance} incorporated a pre-trained audio spectrogram transformer (AST) to robustly predict music genres via transfer learning.}
ExpressiveBailando~\cite{Huang2024EnhancingEI} incorporated frequency information into VQ-VAE and leveraged MERT to add music genre and beat information. Furthermore, TM2D~\cite{Gong_2023_ICCV} built up a shared codebook using both music-to-dance and text-to-motion datasets, enabling dance generation conditioned on both text and music.

The aforementioned methods often relied on complex models and produced low-diversity dance sequences. Recent advances in diffusion models have driven progress~\cite{Qi_2023,EDGE} in music-to-dance generation. EDGE~\cite{EDGE} leveraged a transformer-based diffusion model and the advanced music feature extractor Jukebox~\cite{dhariwal2020jukeboxgenerativemodelmusic} to produce fluid and natural dance movements. 
EDGE greatly influenced subsequent dance generation methods. To ensure that the generated dance was consistent with the music style, DGSDP~\cite{DGSDP} incorporated music style prompts into the diffusion model. To address the issue of motion coherence, LongDanceDiff~\cite{longdancediff} incorporated past movements into the denoising process; BAMD~\cite{BADM}, based on a bidirectional autoregressive diffusion framework, considered both preceding and following frames when generating current frame dance movements. Focusing on the guiding role of lyrics in dance generation, Yin et al.~\cite{LM2D} proposed LM2D, which employed a multimodal diffusion model conditioned on both music and lyrics with consistency distillation. For long dance generation, Lodge~\cite{LODGE} adopted a two-stage coarse-to-fine diffusion framework, using dance primitives as intermediate representations to generate long dances in parallel.

Most of the above methods focused on improving the dance generation network while neglecting the importance of music feature processing. They typically used a frozen Jukebox model or the audio processing toolkit Librosa to extract music features, failing to fully and effectively utilize musical information. As a result, the generated dances often suffer from an insufficient response to musical beats and fail to match the musical style. To address this, BADM incorporated musical beats as input, enhancing the beat alignment of dance generation. M2C~\cite{M2C} demonstrated that the commonly used music features, MFCCs, were highly imbalanced at scale. It proposed a novel normalization process to address this imbalance and used discrete codes as music features for dance generation. In addition to single-person dance generation, recent research has also witnessed a surge in the generation of couple dances~\cite{double1,double2,double4} and group dances~\cite{group1,group2,group3,group4}.

Current music-driven dance generation methods are capable of producing relatively smooth movements and integrating common dance movements. However, there are still challenges such as inconsistency in the overall style of dance, insufficient alignment to musical beats, and lack of coherence in long dances. Future work could focus on enhancing motion coherence, beat alignment, and style consistency, as well as  group and couple dance generation,  hand details,  facial expressions, and  multimodal guidance from lyrics and videos.

\subsubsection{Music-Driven Matched Video} 
Unlike images, videos include temporal dynamics, aligning more naturally with music.
Videos that are semantically and emotionally aligned with music can enhance the expressive potential of musical content by leveraging the intuitive and immediate nature of the visual modality. This task encompasses music-driven video editing and generation\tmp{, as shown in Figure~\ref{fig:m2v}}.
Music-driven video editing selects and crops video clips based on input music to produce music videos (MVs). Audeosynth~\cite{Liao2015audeosynth} extracted optical flow-based video features and music features via note saliency scores, formulating selection and cropping as an optimization problem. It applied Markov chain Monte Carlo sampling to automatically generate video montages using user-input videos. Other works bypassed user-provided videos by constructing dedicated MV corpora and selecting segments that matched the emotion and style of the music, followed by cropping and blending to produce high-quality MVs.
Shin et al.~\cite{Shin} emphasized emotional alignment, using a dimensional emotion model to compute emotional distances between music and video. DJ-MVP~\cite{Fan2016DJMVPAA} leveraged audio similarity and color heuristics to select beat-aligned video segments. Gross et al.~\cite{Gross2019AutomaticRM} employed color histograms, musical key changes, and genre metadata for clip assembly. ApVideor~\cite{Zhang2019AMS} was a music-driven system for fashion video editing by first retrieving music that matched clothing style, then integrating audio and visuals accordingly.

Despite growing interest in music-driven video generation, little research has directly explored the use of music to guide the generation of general video content, which often carries more complex semantic information than music. Music2Video~\cite{Music2Video}, with the assistance of CLIP and Wav2CLIP, leveraged the VQ-GAN method to achieve music video generation guided by both text and music.
In contrast, specific tasks such as singing, instrument playing, and dancing have seen a surge in music-driven video generation methods. Song2Face~\cite{Song2Face}, MusicFace~\cite{MusicFace}, and SINGER~\cite{Li2024SINGER} generated expressive singing videos by aligning facial expressions with vocal tracks. Liu et al.~\cite{liuConducting} proposed the Music Motion Synchronized Generative Adversarial Network (M2S-GAN) to generate orchestral conductor gestures synchronized with music. Liu et al.~\cite{Liu2020BodyMG} and Kao et al.~\cite{Kao_2020} generated 3D body motions for a virtual violinist using RNNs. Similarly, Li et al.~\cite{Li2018SkeletonP} and Bogaers et al.~\cite{Bogaers2020MusicDrivenAG} leveraged LSTMs to synthesize pianist body movements. Yamamoto et al.~\cite{Yamamoto2010GeneratingNH} generated piano-playing hand motions from scores using inverse kinematics, while Shlizerman et al.~\cite{Shlizerman_2018} employed LSTMs to predict full-body and hand gestures of both violinists and pianists.
Unlike previous motion prediction studies, Chen et al.~\cite{Chen2017DeepCA} propose a cross-modal GAN for instrument- and pose-oriented image sequence generation. Sound2Sight~\cite{Cherian2020Sound2SightGV} combines a VAE, GAN, and multimodal transformer to generate drummer performance videos from drum audio and initial frames. Let’s Play Music~\cite{Zhu2020LetsPM} incorporates intra-frame structural features via GCN and inter-frame temporal dynamics via a CNN-GRU, producing high-quality performance videos through a multi-stage STU framework.

Research into music-driven video has mainly focused on music-related scenarios, such as concert video generation and music video editing. This field has been limited by past technological bottlenecks in video processing. However, with recent advances in visual-language models (VLMs) and text-to-video models, we expect more breakthroughs, unlocking greater creativity in music-driven matched visual media.

\section{Music-Oriented Cross-Modal Interaction}
\label{sec:music-oriented}
Music-oriented cross-modal interaction involves music as the output, interacting with visual or textual inputs to retrieve or generate music, enhancing context-aware applications and personalized user experiences. Key tasks include text-driven music retrieval and generation, song generation, controllable music generation, and image and video-driven music retrieval and generation.

\subsection{Text-Driven Matched Music}
\label{subsec:text-guided-music}
Text-guided music matching encompasses two primary subtasks: text-based music retrieval and text-to-music generation. With the development of generation technologies, research has gradually shifted from simple retrieval tasks to more advanced generative models.
\begin{figure*}[!t]
\centering
   \includegraphics[width=1\linewidth]{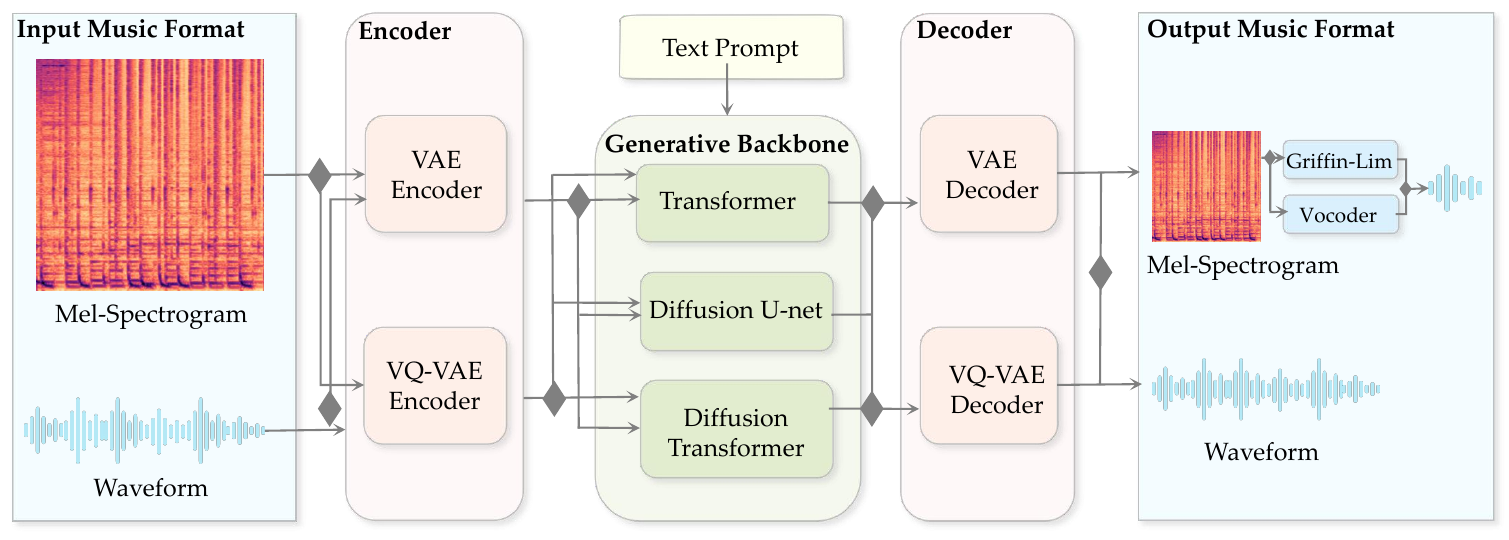}
   \caption{
   Overview of the text-to-music generation pipeline  based on audio representation. Each $\blacklozenge$ represents alternative choices.}

\label{fig:t2m}
\end{figure*}
\subsubsection{Text-Driven Music Retrieval}
\label{par:text-retrieval}

Retrieval tasks are relatively straightforward, with a longer research history. Yu et al.~\cite{Deep2017} explored cross-modal correlation learning, enabling music retrieval using lyrics. Other works~\cite{Multimodal2020, Toward2022, Musical2024} have employed multimodal metric learning, contrastive learning, and the design of music-specific word embeddings to develop more general and high-performing text-to-music retrieval systems, contributing high-quality datasets and  comprehensive evaluation benchmarks. Earlier text-to-music retrieval efforts were constrained by fixed vocabularies, lacking the flexibility of natural language. Doh et al.~\cite{Enriching2024} addressed this by utilizing large language models to generate rich music descriptions, enabling joint training across multiple datasets. These systems relied on audio representations. In contrast, CLAMP~\cite{wu2023clamp} stood out as a contrastive learning and pretraining-based cross-modal symbolic music information retrieval model, supporting symbolic semantic search and zero-shot classification.

\subsubsection{Text-to-Music Generation}

Text-to-image generation has made significant advances and garnered considerable attention in recent years. Using natural language to guide image creation allows individuals without drawing skills to produce artwork, as seen in the AI-generated  Space Opera Theatre~\cite{roose2022ai}. Music, another art form with a high entry barrier, has similarly drawn researchers to explore text-to-music generation. This approach lowers the barriers to music creation and acts as a bridge in multimodal data, enabling richer multimodal music tasks. Most works focus on using text to control high-level musical features for generation, categorized by the data representation used: symbolic representation or audio representation.

Text-to-symbolic music generation refers to guiding a model with text to generate structured symbolic music data, such as MIDI, REMI, and ABC notation. MuseCoco used a two-stage framework, first extracting musical attributes (e.g., rhythm, key) from text, then generating music based on these attributes to enhance musicality and controllability. MeloTrans~\cite{wang2024melotrans} combined neural networks with human compositional principles, particularly motif development rules, to improve the structure and musicality of symbolic music generation.

Several studies have explored the use of LLMs for symbolic music generation. ChatMusician treated music as a second language, pre-training and fine-tuning LLaMA2~\cite{touvron2023llama2openfoundation} on ABC notation to understand and generate music without external multi-modal structures. MuPT introduced the SMT-ABC symbolic representation and the symbolic music scaling law, revealing the relationship between training data scale and model performance.

In recent years, the rapid development of large-scale generative models has significantly advanced text-to-music generation based on audio representations. These approaches can be broadly classified into Transformer-based autoregressive methods and non-autoregressive methods, including diffusion model-based approaches, as shown in Figure~\ref{fig:t2m}.
Autoregressive methods typically encode music into discrete tokens and employ Transformer language models as the generative backbone.
MusicLM~\cite{MusicLM2023}, built upon AudioLM~\cite{audiolm2022}, incorporated textual control via MuLan, using a semantic-acoustic multi-stage sequence modeling approach to generate minutes-long, coherent music. 
MusicGen~\cite{copet2024simple}, based on AudioGen~\cite{audiogen2023}, introduced an efficient token-interleaving pattern, enabling high-quality single-stage mono or stereo music generation. \tmp{It decomposes audio into $K$ parallel codebook streams via residual vector quantization (RVQ). These streams are interleaved into a single sequence $\mathbf{X}$ to be modeled autoregressively:$$P(\mathbf{X}) = \prod_{i} P(x_i | x_{<i}, \mathbf{C}), $$where $\mathbf{C}$ is the conditioning context. This pattern enables the model to effectively capture dependencies across different codebook levels within a single stage.}
A significant drawback of autoregressive models is their slow inference speed. To address this, VampNet~\cite{VampNet2023}, STEMGEN~\cite{stemgen2023}, and MAGNET~\cite{magnet2024} utilized non-autoregressive Transformer models with masked modeling, achieving efficient text-to-music generation. MusicRL~\cite{musicrl2024} was the first music generation system fine-tuned with human feedback. Building on MusicLM, it employed reinforcement learning (RL) with multiple reward functions and user preference data, enhancing generation quality and alignment with human preferences. 
\tmp{To optimize for diverse objectives, MusicRL utilizes a weighted composition of reward functions defined as:$$R(c, a) = \sum_{k \in \mathcal{K}} w_k R_k(c, a), $$where $c$ and $a$ represent the text context and generated audio, and $w_k$ is the weight for each individual reward component $R_k$ (such as text-alignment, audio quality, and human preference). This multi-objective approach allows the model to simultaneously improve both technical fidelity and subjective appeal.}

Diffusion models are another prevailing non-autoregressive generative framework. Riffusion~\cite{Forsgren_Martiros_2022} fine-tuned the text-to-image model Stable Diffusion (SD) using a small dataset of audio spectrograms paired with text descriptions. Subsequent works, including Make-an-Audio~\cite{make2023}, AudioLDM~\cite{audioldm2023}, AudioLDM2~\cite{audioldm2}, \tmp{Tango~\cite{ghosal2023tango}, Tango2~\cite{majumder2024tango2}, Diff-a-riff~\cite{nistal2024diffariff} and Auffusion~\cite{xue2024auffusion}} leveraged diffusion models to generate text-guided audio content, spanning speech, sound, and music. While these methods advanced text-to-audio generation, their music-specific capabilities lagged behind dedicated text-to-music models. Dedicated models like MusicLDM~\cite{musicldm2023}, Noise2Music~\cite{Noise2Music2023}, and Gen1~\cite{gen1_2024} employed diffusion models with 1D or 2D UNet architectures to model music waveforms or spectrograms, enabling text-guided music generation of fixed durations. Later efforts focused on refining diffusion-based text-to-music models to improve quality, flexibility, and efficiency. Stable Audio~\cite{SD2024} introduced temporal embedding conditions to flexibly control the duration of generated music. MeLoDy~\cite{lam2024efficient}, adapted from MusicLM, retained semantic token modeling with a language model while using a dual-path diffusion model for acoustic tokens, ensuring quality and enabling efficient music generation.

The DiT (Diffusion Transformer) model~\cite{peebles2023scalable}, excelling in text-to-video generation (e.g., Sora), inspired further innovation in generative frameworks, with some works adapting it for text-to-music generation. Stable~Audio~2~\cite{evans2024stable} replaced the UNet-based diffusion model in Stable Audio by a Transformer-based architecture, enabling longer-duration, higher-quality music generation, alongside creative capabilities like audio style transfer. 
Although diffusion models excel at generating high-quality content, they suffer from numerous sampling steps and prolonged inference times. Flow matching techniques, which directly learn transformations between distributions, typically offer faster inference and fewer sampling steps. MusicFlow~\cite{musicflow2024} employed cascaded semantic and acoustic flow matching models, achieving competitive results with significantly fewer model parameters while boosting inference speed, thus improving efficiency. FluxMusic~\cite{flux2024} integrated rectified flow matching into the DiT model for text-to-music generation, using a joint text-music dual-flow and music single-flow architecture. This combined the high generation quality of DiT with the efficiency of flow matching. Other works, such as ConsistencyTTA~\cite{ConsistencyTTA2024}, Presto~\cite{presto2024}, and DITTO-2~\cite{ditto2}, achieved accelerated music generation through targeted distillation.

\tmp{Previous text-to-music works were limited to generating instrumental music, but recent advanced studies have begun exploring text-to-song generation. Industry tools such as Suno \url{https://suno.com/blog/introducing-v4-5}, Udio \url{https://www.udio.com/blog/introducing-v1-5}, Seed-Music~\citep{seed2024}, and Meruka\ \url{https://www.mureka.ai} have demonstrated remarkable capabilities in generating high-fidelity songs with complex structures. Academic efforts have followed closely, with the development of language model-based song generation approaches.
Melodist~\citep{hong2024text} first defined the text-to-song synthesis task, which integrated singing voice synthesis and vocal-to-accompaniment generation into a unified framework. MelodyLM~\citep{li2024accompanied} achieved high-quality, fully text-controlled melody generation by integrating language-model-based vocal synthesis with latent-diffusion-based accompaniment, offering maximum creative flexibility with minimal user requirements. Songcreator~\citep{lei2024songcreator} employed a dual-sequence language model to capture the intricate dependencies between vocals and accompaniment, enabling a more versatile and controllable end-to-end song generation and editing framework. YuE~\citep{yuan2025yue} scaled the LLaMA2 architecture to music generation by introducing a track-decoupled next-token prediction strategy and structural progressive conditioning, enabling the synthesis of full-length songs with high lyrical alignment and sophisticated vocal-accompaniment coordination. SongGen~\citep{liu2025songgen} used a single-stage auto-regressive transformer for controllable song generation, which streamlined the synthesis process into a unified framework to mitigate error accumulation; it supported dual-track output modes for flexible downstream applications. LeVo~\citep{lei2025levo} introduced a dual-token modeling framework that concurrently processed mixed and dual-track representations to optimize vocal-instrument harmony, while further enhancing musicality and instruction-following through a multi-preference alignment strategy based on direct preference optimization (DPO). Subsequently, Lam et al.~\cite{lam2025analyzable} introduced MusiCoT, a CoT prompting technique for music generation, achieving high-fidelity music generation in two stages and opening new avenues for future exploration. Advances in diffusion models have further propelled this field. For instance, DiffRhythm~\citep{ning2025diffrhythm} is a simple yet scalable latent diffusion-based model that enables the rapid synthesis of full-length songs, overcoming the inference speed and duration limitations of previous multi-stage or autoregressive approaches. ACE-Step~\citep{gong2025ace} further improved upon DiffRhythm by incorporating song structure understanding.}

Text-to-music generation serves as the foundation of multimodal music generation, continuously pushing the boundaries of a model's ability to generate music. This evolution spans from symbolic music (e.g., MIDI representations) to audio-based music, and from instrumental music to songs, with generated music having greater realism, flexibility, and diversity. Although current models generate high-quality songs and some support audio prompts, they still face several challenges, including limited fine-grained control (e.g., of BPM and style), imprecise temporal melody control and weak long-term musical structure. Future improvements may be achieved by collecting higher-quality and more diverse datasets, exploring generation framework designs tailored to music, and incorporating domain-specific knowledge from music composition.

\subsection{Music Generation with Fine-grained Control}
\tmp{Music generation with fine-grained control aims to produce music that adheres to specific requirements beyond text prompts by utilizing multiple conditions. Going beyond the high-level semantic control in text-to-music generation, this task enforces control over specific musical attributes such as chords, rhythm, and melody.}

\tmp{Several approaches have achieved fine-grained control by explicitly integrating musical prompts into the generation process. For instance, Mustango~\cite{melechovsky2023mustango} expanded the scope of text-driven generation by utilizing rich captions that include specific instructions related to chords, beats, tempo, and key.  
Similarly, in the symbolic domain, MuseBarControl~\cite{shu2024musebarcontrol} enhanced the control prompts of MuseCoco by introducing bar-level attribute tags and positional embeddings.
To achieve precise temporal control, researchers have developed advanced adapter mechanisms.} Music ControlNet~\cite{wu2024music} adopted a strategy similar to ControlNet~\cite{zhang2023adding}, achieving temporal control over melody, dynamics, and rhythm for the first time. JASCO~\cite{tal2024joint} used a flow-matching paradigm and \tmp{low-dimensional bottleneck projections} to achieve temporally controllable text-to-music generation, offering superior melody adherence, accuracy, and local control. MusiConGen~\cite{lan2024musicongen} employed an in-attention mechanism and designed an efficient fine-tuning strategy to control rhythm and chords in music. \tmp{Targeting the complex task of cover song generation, SongEcho~\cite{li2026songecho} employed an instance-adaptive linear modulation method to jointly synthesize new vocals and accompaniment while preserving the original melody.
Unlike the aforementioned supervised methods, DITTO~\cite{DBLP:conf/icml/NovackMBB24} explored training-free control. It formulated the task as optimizing the initial noise latent vector during inference, enabling pretrained models to accept arbitrary control signals without additional fine-tuning. However, the performance of such direct inference-time optimization remains limited.}

\tmp{Music generation with fine-grained control} has advanced by combining new techniques with inputs like musical attributes and textual information to better meet creators' intentions, thereby enhancing realism, flexibility, and fine-grained control over rhythm, melody, and dynamics. Future progress may involve collaborative human–AI refinement systems.

\begin{figure*}[!t]
\centering
   \includegraphics[width=1\linewidth]{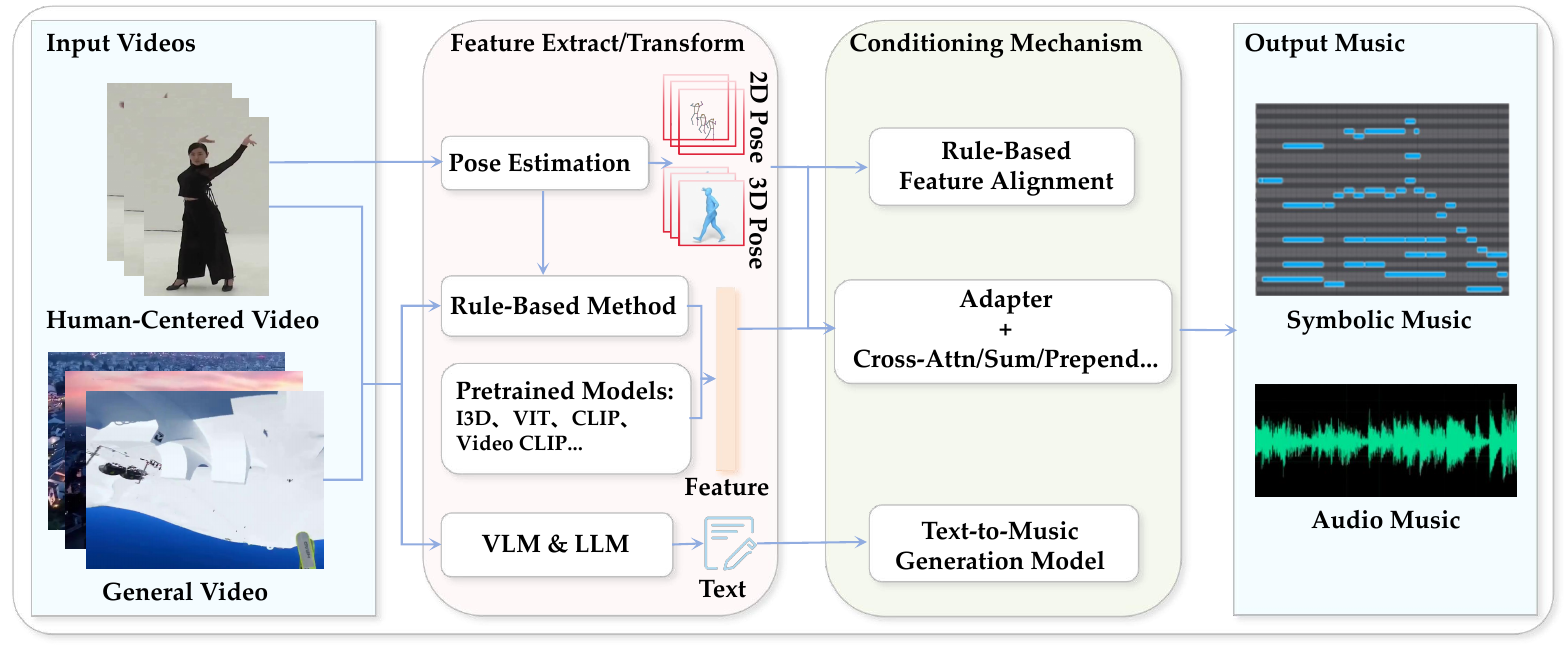}
   \caption{
   Overview of the video-to-music generation pipeline based on audio representation. }

\label{fig:v2m}
\end{figure*}

\subsection{Image-Driven Matched Music}

\subsubsection{Image-Driven Music Retrieval} 
Image-driven music retrieval aims to automatically match or recommend music based on image content, such as scenes, emotions, or artistic styles, enhancing the user experience in multimedia interactions and personalized recommendations.

Hsia et al.~\cite{hsia2018representation} proposed a representation learning framework to bridge the gap between images and music, extracting high-level semantic features and constructing a unified embedding space for image-to-song retrieval. 
Emotion served as a bridge between images and music, quantifying their similarity in the continuous valence-arousal (VA) space. Sasaki et al.~\cite{sasaki2013affective} mapped images and music to a VA emotional plane using visual features and acoustic analysis, recommending music by calculating their distance in the plane.  Zhao et al.~\cite {zhao2020emotion} utilized cross-modal deep continuous metric learning to model the shared embedding space while preserving cross-modal similarity with unimodal VA relationships, significantly improving the finesse of emotion matching. Baijal et al.~\cite{baijal2021analyzing} further incorporated style into the retrieval process, recommending music for artworks based on both style and emotion, and for photographs, based solely on emotion. Diff4Steer~\cite{bao2024diff4steer} generated multi-directional music discovery paths from user queries by synthesizing diverse seed embeddings through a lightweight diffusion model. It supported image or text-guided retrieval, combined with nearest neighbor search for flexible and controllable music retrieval.

\subsubsection{Image-to-Music Generation}

Image-to-music generation transforms visual information into musical sequences that match the semantic or emotional expression of images. The main challenge lies in establishing subjective cross-modal associations while maintaining emotional and artistic consistency of music.  Current research focuses on emotion mapping, cross-modal alignment, and generative model optimization, resulting in diverse approaches.

Jukepix~\cite{wang2018jukepix} proposed a cross-modality model that converted impressionist paintings into multitrack music using deep convolutional generative adversarial networks (DCGANs). Mendis et al.~\cite{mendis2022show} combined a CNN with LSTM to generate melodies matching the emotional tone of images, though only monophonic melodies were supported, and musical instruments and dynamics were not considered. Some works have used the VA emotional space to align music with the emotional tone of images. Wang et al.~\cite{wang2023continuous} enhanced emotional consistency by establishing a mapping  between images and music in a continuous emotional space. Kundu et al.~\cite{kundu2024emotion} introduced a CNN-Transformer architecture that combined a pre-trained CNN image feature extractor with Transformer to generate musically and emotionally consistent MIDI sequences. They also introduced a VA loss function that directly optimized emotion alignment, avoiding the complexity of contrastive learning. With the advance of LLMs, Hisariya et al.~\cite{hisariya2024bridging} proposed a two-stage framework, where images were first converted into emotional descriptions and then music was generated using a fine-tuned MusicGen model.

With the rise of diffusion models, researchers have explored their use for cross-modal fusion. MELFUSION~\cite{melfusion2024} introduced a music diffusion model with a novel `visual synapse' that effectively infused image semantics into a text-to-music model. Zhao et al.~\cite{zhao2024cptgz} combined latent diffusion with a Transformer to automatically generate waveform guzheng music from Chinese paintings.

Image-driven music matching has evolved along two main trajectories. Music retrieval bridges visual and auditory modalities through semantic and emotional alignment. In contrast, image-to-music generation has progressed from early GAN- and LSTM-based models generating simple monophonic melodies to more advanced approaches emphasizing emotional consistency and using diffusion models for high-fidelity, culturally specific outputs. 
However, retrieval systems struggle with subjective emotional interpretation and dataset biases, while generative models face challenges in maintaining long-term musical coherence. Current evaluation methods rely on synthetic datasets or limited cultural contexts, constraining real-world applicability. Future advances may arise from integrating large multimodal models, iterative human feedback, and more diverse datasets to enhance creativity and utility in image-driven matched music.

\subsection{Video-Driven Matched Music}
\label{subsec:video-guided-music}

Music and videos complement each other: music enhances video atmosphere and emotion, while videos broaden dissemination and perceptual dimensions of music. This synergy drives video-guided matched music tasks, including (i) video-guided music retrieval, which uses video features to find matching music from collections, and (ii) video-guided music generation, which creates tailored music using video features via generative models such as diffusion models and autoregressive models, as shown in Figure~\ref{fig:v2m}.

\subsubsection{Video-Driven Music Retrieval}
\label{par:video-retrieval}

Early work, such as Sasaki et al.~\cite{affective2015}, explored emotion analysis of video and audio features and realized emotion-consistent music recommendation using a non-deep learning approach based on V-A plane distances. Subsequently, CBMVR~\cite{CBVMR2018} employed neural networks to embed video and music features into a shared space and achieved bidirectional music-video recommendation using cross-modal ranking constraints and soft intra-modal structural constraints. Suris et al.~\cite{suris2018cross} improved this by extracting audio features with neural networks instead of computing statistical handcrafted features and optimized recommendation with similarity and classification losses. Li et al.~\cite{li2019query} introduced emotion label constraints into a dual-stream network and achieved efficient cross-modal music retrieval. Prétet~\cite{cross2021} built on CBMVR and explored design choices for music-video cross-modal recommendation, including optimization of audio features and loss functions.
CMVAE~\cite{corssmodal2021} used a VAE model to learn associations between music and video through cross-modal reconstruction constraints, enhancing music retrieval with video and text.
Earlier works overlooked temporal alignment and artistic correspondence in audio-video cross-modal recommendation.
Suris et al.~\cite{itstime2022} used a Transformer to model relationships between visual features and music tag embeddings, achieving self-supervised audio-video recommendation via contrastive learning.
Seg-VM-net~\cite{v2mr2023} also addressed temporal alignment issues.
The above methods neglected noise in data under the subjective task of video-music pairing. 
SSVMR~\cite{cheng2023ssvmr} handled label noise, enhanced key video segment capture, and improved model robustness through semi-supervised self-training, saliency-based span mixing, and reverse retrieval. 
Considering music selection bias in user-created content due to historical preferences, Yi et al.~\cite{deconfounded2024} designed a  video background music recommendation algorithm to combat the problem and increase recommendation system accuracy. More flexibly, ViML~\cite{mckee2023language} retrieved music matching video content and stylistic descriptions by integrating representations of videos and texts.

\subsubsection{Video-to-Music Generation}
\label{par:video-generation}
Video-to-music generation initially focused on symbolic representation-based methods, where generating music for performance videos first attracted researchers. Sight to Sound~\cite {koepke2020sight} employed a ResNet model to predict pitch onset events from video frames, aiming to transcribe piano performance videos into MIDI data, mitigating issues like polyphonic instruments and noise interference encountered in audio-to-MIDI transcription. Audeo~\cite{audeo2020} introduced a two-stage ResNet-GAN network to generate music from silent piano videos, emphasizing audio reconstruction. The same authors later proposed MI Net~\cite{minet2020}, which leveraged VQ-VAE to directly generate music waveforms for piano performances. Foley Music~\cite{gan2020foley} extended video-to-music generation from piano to multiple instruments using human keypoints. It used a Graph-Transformer framework for better generality, accuracy, and editing flexibility. Recent works~\cite{wangfrieren, wang2024v2a} have also generated music from video but prioritized semantic and motion alignment over MIDI-level accuracy.

Dance, as an art form highly related to music, has led to ongoing exploration of the generation of music for dance videos. Dance2MIDI~\cite{han2024dance2midi} and DanceComposer~\cite{liang2024dancecomposer} generated rhythm and style-consistent multi-track symbolic music from dance videos, achieving rich symbolic music generation. Dance2MIDI additionally constructed a paired dataset DMIDI of dance and multi-instrumental music. In contrast, D2M-GAN generated audio waveforms directly using VQ representations, I3D, and motion features with GANs, and compiled the TikTok wild dance music dataset. CDCD~\cite{zhudiscrete}, LORIS~\cite{yu2023long}, and MDM~\cite{mdm2023} used human motion information (2D keypoints or 3D motion features) as conditions for diffusion models to generate audio music  aligned with dance rhythms. CDCD and LORIS also employed I3D features as additional condition controls. CDCD supported broader cross-modal generation tasks such as text-to-image generation. The direct audio generation works mentioned above all required training models from scratch, resulting in low-quality music due to the limited availability of public dance datasets. Li et al.~\cite{li2024dance2music} enhanced text-to-music models for dance video scoring using encoder-based text inversion methods. MotionComposer~\cite{wang2025motioncomposer} improved synchronization and style consistency between rhythm and motion videos through retrieval-augmented generation (RAG) technology and progressive training strategies. 

In addition to the aforementioned human-centric videos, the internet contains plenty of  video-music data pairs with diverse scenes, making general video background music generation an increasingly prominent research area. 
CMT~\cite{cmt2021} employed a rule-based approach to align video motion features with corresponding attributes of symbolic music, achieving video background music generation in the absence of paired video-music datasets. Video2Music~\cite{kang2024video2music} extracted semantic, motion, and emotional features from videos, encoded them using a Transformer encoder, and subsequently input them into a Transformer decoder to autoregressively generate MIDI notes. 
V-MusProd~\cite{V-MusProd2023} presented a multi-stage Transformer framework for video background music generation.
Diff-BGM leveraged visual features and textual descriptions from videos to guide a latent diffusion model to generate piano rolls with a segment-aware cross-attention module.
Nevertheless, symbolic music-based approaches faced data-scale issues and often generated less expressive and diverse outputs.

Recent studies have explored the direct generation of music waveforms from video. V2Meow~\cite{v2meow2023} used a multi-stage autoregressive model to generate visually aligned music from multidimensional video features, allowing high-level feature control via text. MuMu-LLaMA, a multimodal music understanding and generation model, leveraged a large language model as a bridge, building on text-to-music models to enable video-to-music generation, utilizing the MUVideo dataset focused on instrumental performances. VidMuse~\cite{vidmuse2024} constructed a large-scale dataset of 360,000 video-music pairs and employed long-short-term visual modules to generate music synchronized with video.
S2L2-V2M~\cite{kai2024video}, a novel framework, leveraged a memory-augmented state space model to enable efficient and effective long-term video-to-music generation using large language models.
M2M-Gen~\cite{sharma2024m2m}, VEH~\cite{tong2024video} and SONIQUE~\cite{zhang2025sonique} explored the use of text as a bridge to generate background music for videos, avoiding the reliance on paired video and music data.
These approaches prioritized aligning video semantics with music but lacked fine-grained rhythm control.  
VMAS~\cite{vmas2024} refined the standard cross-entropy loss to focus on video-music alignment by detecting video and music beats. MuVi~\cite{muvi2024}, based on a DiT model with a flow-matching objective, achieved semantic alignment and rhythmic synchronization in video-to-music generation through a visual adapter. 
\tmp{VeM~\cite{tong2025video} introduced a hierarchical `music conductor' strategy within a diffusion framework to strictly enforce semantic and rhythmic synchronization via storyboard-guided attention.}
VidMusician~\cite{vidmusician2024} and GVMGen~\cite{zuo2025gvmgen} compiled diverse paired video-music datasets and developed video background music generation methods with rhythm and semantic alignment based on MusicGen.
MuVi, VidMusician, and GVMGen harnessed the generative capabilities of pretrained music generation models, with MuVi and VidMusician incorporating control conditions via token-level addition, which aided rhythm alignment. 
More specifically, FilmComposer~\cite{xie2025filmcomposer} and HPM~\cite{qi2024harmonizing} respectively employed large model-driven agent collaboration and designed the Film Score ControlNet to achieve controllable generation of film background music. 
Video2Song~\cite{yin2025video2song} harnessed retrieval-augmented VLMs and LLMs to generate structured, video-aligned prompts for music generation, functioning as a proto-agent that exhibited rudimentary agentic behaviors.
Furthermore, several recent video generation studies~\cite{kong2024hunyuanvideo, polyak2025moviegencastmedia} have integrated background music generation as a subtask.

Video-to-music generation has great potential but is currently limited to instrumental music. It faces challenges like insufficient high-quality datasets, weak music-video correspondence, and a lack of fine-grained rhythm alignment. To enhance video-to-music generation, future work should focus on collecting diverse, high-quality music-video datasets, refining audio-video feature alignment, and developing reasoning-capable generative models, thereby improving quality and flexibility, meeting user needs, and advancing commercialization.

\tmp{\subsection{Singing Voice Generation}}
\subsubsection{Singing Voice Synthesis (SVS)} 

SVS can be viewed as an extension of the speech synthesis task, aiming to generate natural, expressive, and high-fidelity singing voices based on musical scores and lyrics. Current deep learning-driven SVS systems primarily adopt an acoustic model with vocoder architecture similar to speech synthesis. 
In fact, numerous SVS systems utilize modified versions of speech synthesis models.
Despite significant advances in two-stage human voice generation, training the neural acoustic model and vocoder separately can cause a mismatch between training and inference, leading to suboptimal performance. Recent research has shifted towards exploring end-to-end unified frameworks to improve synthesis quality.

ByteSing~\cite{gu2021bytesing} enhanced acoustic modeling by combining a duration-allocated encoder–decoder structure with powerful WaveRNN vocoders, achieving significant control over pitch and spectral fidelity in the generated singing voice. XiaoiceSing~\cite{lu2020xiaoicesing} followed the main architecture of FastSpeech~\cite{ren2019fastspeech} while incorporating musical features, improving fundamental frequency prediction through residual connections, and designing syllable duration loss to enhance rhythm modeling.

GANs have been widely explored for SVS. WGANSing~\cite{chandna2019wgansing} proposed a block-wise acoustic modeling approach using vocoder features to decouple pitch and timbre, enabling temporal dependency modeling and smooth synthesis. HiFiSinger~\cite{chen2020hifisinger} addressed high sampling rates by proposing a sub-frequency GAN and a multi-length GAN to optimize frequency bandwidth and long sequence modeling, introducing fundamental frequency ($F_0$) and $V/UV$ as acoustic features for high-fidelity SVS. SingGAN~\cite{huang2022singgan} designed source excitation modules and global-local discriminators to optimize continuity, high-frequency reconstruction and low-frequency details. XiaoiceSing2~\cite{wang2022xiaoicesing} employed a GAN architecture with a FastSpeech-based generator and multi-band discriminator to address mel-spectrogram smoothing issues observed in XiaoiceSing.

DiffSinger pioneered the use of diffusion models in SVS, generating mel-spectrograms through iterative Markov chain denoising. \tmp{To enhance inference efficiency and detail restoration, it proposed a shallow diffusion mechanism. Instead of starting from Gaussian noise $x_T$, the reverse process begins at a boundary time-step $K$ ($K \ll T$), where the initial latent $x_K$ is constructed from an auxiliary predictor’s output $x_\mathrm{aux}$ and a noise term $\epsilon$:$$x_K = \sqrt{\bar{\alpha}_K} x_\mathrm{aux} + \sqrt{1 - \bar{\alpha}_K} \epsilon.$$ By refining the coarse spectrogram $x_\mathrm{aux}$ via $K$ iterations of the denoising function $\epsilon_\theta(x_t, t, \mathcal{C})$, where $\mathcal{C}$ represents musical conditions, DiffSinger effectively mitigated the over-smoothing issues of traditional GANs while significantly accelerating the sampling process.} CoMoSpeech~\cite{ye2023comospeech} applied a consistency constraint to distill a consistency model from a well-designed diffusion-based teacher model, requiring only a single sampling step to generate high-quality audio.

End-to-end models simplify traditional cascaded workflows, improving synthesis efficiency and consistency. Jukebox used a multiscale VQ-VAE to compress raw audio to discrete codes and modeled them using autoregressive Transformers. VISinger~\cite{zhang2022visinger} followed the VITS~\cite{kim2021conditional} architecture, and was enhanced with a length regulator, frame prior network, and modified duration predictor. VISinger 2~\cite{zhang2023visinger2} integrated a digital signal processing (DSP) synthesizer into the decoder to address text-to-phase mapping, glitch artifacts, and low sampling rate issues. \tmp{Muse-SVS~\cite{kim2023muse} introduced a joint embedding space that uniformly represents pitch, energy, and phoneme duration, allowing precise control over attribute variations and accurate emotional expression with different intensities. Prompt-Singer~\cite{wang2024prompt} enabled natural language control over singer gender and vocal range by decoupling pitch from melody representation, advancing human-AI interactive music creation.} 

Despite significant advances, current systems still face challenges in generating expressive results, real-time controllability, and producing coherent long-form outputs with natural characteristics. Future improvements may arise from multimodal learning approaches, self-supervised techniques to address data scarcity, and integration of music theory. 

\subsubsection{Singing Voice Conversion (SVC)}

SVC aims to transform the timbre, singing techniques, or style of a source singer into those of a target singer  while preserving the song content and melody. Luo et al.~\cite{luo2020singing} proposed a dual-encoder framework that separately modeled latent representations of singer identity and singing techniques, enabling many-to-many conversion through vector operations. Lu et al.~\cite{lu2020vaw} introduced a VAW-GAN-based framework that disentangled singer identity and prosody from phonetic content to generate spectral features for unseen target singers with improved $F_0$ rendering. Li et al.~\cite{li2021ppg} proposed an end-to-end architecture with dual encoders for content and acoustic features, enhanced by adversarial and regressive modules to improve timbre and melody modeling. FastSVC~\cite{liu2021fastsvc} utilized a conformer-based phoneme recognizer and a lightweight waveform generator to achieve high conversion performance with fast inference speed. Jayashankar et al.~\cite{jayashankar2023self} circumvented disentanglement training and utilized ASR-finetuned self-supervised features as inputs to a HiFi-GAN vocoder. Ning et al.~\cite{ning2023vits} proposed a VITS-based singing voice conversion model using Whisper-extracted bottleneck features and multi-scale pitch information, trained with a three-stage strategy to adapt it to limited target speaker data.

The emergence of diffusion models has brought breakthrough advances in SVC. DiffSVC~\cite{liu2021diffsvc} pioneered the application of denoising diffusion probabilistic models to SVC tasks, achieving high-quality spectrogram generation by predicting Gaussian noise and surpassing traditional methods in naturalness and voice similarity. However, its iterative sampling process limited inference speed. To address this, CoMoSVC~\cite{lu2024comosvc} achieved single-step sampling through consistency model distillation, improving inference speed while maintaining high-quality generation. Similarly, LCM-SVC~\cite{chen2024lcm} reduced the inference steps of latent diffusion models (LDMs) using latent consistency distillation techniques, significantly accelerating the generation process. To address the issue of timbre leakage, LDM-SVC~\cite{chen2024ldm} pretrained a variational autoencoder based on the open-source So-VITS-SVC project and introduced a singer-guided training method to further suppress the timbre of the original singer. Additionally, the visual analysis system SingVisio~\cite{xue2024singvisio} enhanced model generation mechanism interpretability through interactive diffusion process visualization.

SVC has evolved from early approaches leveraging disentangled representations via VAEs and adversarial training to advanced diffusion-based frameworks, achieving high-fidelity timbre transformation while preserving melody and content. Despite progress, challenges persist in fine-grained disentanglement of timbre, pitch, and expressive nuances, particularly in cross-linguistic or low-resource scenarios. Future directions may focus on improved disentanglement mechanisms, lightweight architectures, and standardized benchmarks.

\subsection{Multimodal-Driven Matched Music}
\label{subsec:multimodal-driven-music}
The foregoing discussion outlined research on acquiring matched music through methods driven by various single modalities. In contrast, other studies aim to develop a unified model that leverages multiple modalities for collaborative music acquisition driven by diverse inputs.
Shen et al.~\cite{shen2020enhancing} enhanced music recommendation by constructing a multimodal dataset, defining handcrafted and deep features, and proposing an attention-based multimodal autoencoder (AMAE) that combined text and visual features.
However, generating music rather than retrieving it represents a prevailing trend in current developments.
MuMu-LLaMA employed a large language model as a bridge between multiple pretrained modality encoders and a music generator, achieving the generation of music conditioned on text, images, and video. Mozart’s Touch~\cite{xu2024mozart} utilized BLIP~\cite{pmlr-v162-li22n} to convert visual content into  text, directly expanding the multimodal music generation capabilities of the text-to-music model MusicGen. VMB~\cite{rag2024}, built upon Stable Audio, demonstrated superior performance in multimodal music generation tasks through dual-track retrieval augmentation. More comprehensively, AudioX~\cite{tian2025audiox}, a unified diffusion transformer, generated high-quality audio and music from diverse multimodal inputs using a novel masked training strategy.
Instead of directly generating audio, XMusic~\cite{tian2025xmusic} introduced a novel framework that leveraged flexible multimodal prompts to produce generalized and emotionally controllable symbolic music.

Works on arbitrary-to-arbitrary generation have treated multimodal music generation as a subtask.
CoDi~\cite{tang2024any} used multiple modality encoders and diffusion models with alignment to generate arbitrary output combinations from varied inputs, but struggled with complex tasks. NExT-GPT~\cite{nextgpt2024}, an end-to-end multimodal LLM, integrated components with alignment learning and instruction tuning for robust input-to-output generation and strong complex task handling. AnyGPT~\cite{zhan2024anygpt} employed a unified multimodal tokenizer, converting diverse modalities (images, music, speech) into discrete semantic tokens for uniform processing and generation in an LLM.

While exhibiting basic capabilities in generating matched music, multimodal music generation offers a promising, flexible cross-modal interaction system between music and multimodal data, with significant potential for enhancing user music creation experiences.
However, integrating multiple modalities poses modeling challenges, with room for improvement in deep multimodal alignment and understanding, adaptability, and personalized generation capabilities.

\section{Bidirectional Music Cross-Modal Interaction}
\label{sec:bidirectional-music}
Bidirectional cross-modal music interaction involves the dynamic interplay between music and other modalities, enabling both the generation of music from diverse inputs and the extraction and conversion of music information into other  content modalities. Specifically, it takes the following forms:
\begin{enumerate}
\item \emph{Multimodal music generation and understanding:} a model generates music from visual, textual, or other inputs and extracts information from music for understanding.
\item \emph{Multimodal music editing:} editing music using other modalities as conditions.
\item \emph{Music agents:} enabling deep, multimodal, multi-turn dialogues for personalized, context-aware music creation and interaction.
\end{enumerate}

\subsection{Multimodal Music Understanding and Generation}
Some studies~\cite{liang2024pianobart} have focused on symbolic music generation and understanding. The ABC notation shares similarities with text in terms of data format. ChatMusician, based on LLaMA2, underwent continual pretraining and fine-tuning on music data represented in ABC notation. It treated music as a second language, achieving strong performance in music understanding, generation, and linguistic capabilities. However, it faced challenges such as hallucination and memory degradation.
Other studies~\cite{huang2024audiogpt} have focused on audio-based music understanding and generation. MuMu-LLaMA integrated multiple pretrained modality feature encoders via designed adapters, unifying different modality information into a LLaMA model. Multimodal music understanding and generation were achieved by organizing multimodal datasets and designing multi-stage training strategies.

\subsection{Multimodal Music Editing}
Some autoregressive text-to-music generation models~\cite{MusicLM2023,copet2024simple} have supported melody-guided generation. 
Hou et al.~\cite{hou2025editing} enhanced the DiT model using ControlNet, introducing top-$k$ CQT as a melody prompt to achieve the above task.

For comprehensive music editing, AUDIT~\cite{wang2023audit} and InstructME~\cite{InstructMe2023}, both based on diffusion models, were trained from scratch to implement audio editing guided by human instructions. They supported various audio editing tasks, including addition, deletion, replacement, restoration, and super-resolution. InstructME further supported mixing operations. 

Several works have explored fine-tuning strategies for music editing.
Instruct-MusicGen~\cite{instruct2024} incorporated an audio fusion module and a text fusion module into MusicGen, fine-tuning it with instructions to achieve music editing.
Lin et al.~\cite{content2023,arrange2024} enhanced content-controllable music generation by designing parameter-efficient fine-tuning (PEFT) methods for MusicGen. 
AP-Adapter~\cite{adapter2024} incorporated audio as an input for AudioLDM2, which extracted audio features using AudioMAE and integrated them into the model through an adapter, to achieve music editing.
Some studies have explored training-free strategies for music editing.
Manor et al.~\cite{zero2024} proposed two zero-shot audio editing methods: Zero-shot Text-to-Audio (ZETA) editing and Zero-shot Unsupervised (ZEUS) editing. 
Zhang et al.~\cite{musicmagus2024} determined  editing direction by computing the difference between two embedding sets of different captions.
MELODYFLOW~\cite{lan2024high} introduced a flow-matching architecture with regularized latent inversion, significantly improving stereo generation fidelity and consistency in text-guided editing. MEDIC~\cite{liu2024medic} proposed a disentangled inversion control framework using a triple-branch architecture and coordinated attention to enable zero-shot music editing.
Additionally, some studies have focused on music style transfer. Li et al.~\cite{li2024music} and Kim et al.~\cite{traininfree2024} achieved music style transfer through textual inversion and training-free approaches, respectively.

\subsection{Music Agents}
\tmp{Music agents mark a new era in AI music systems, leveraging large language models (LLMs) as a central brain to bridge multimodal perception and generation through reasoning and planning.}
AudioGPT combined audio foundation models with a large language model, enabling music understanding and generation via multi-turn dialogue, though limited by prompt engineering, token constraints, and model performance. Loop Copilot~\cite{zhang2023loop} and MusicAgent~\cite{yu2023musicagent} used LLMs to integrate music tools, automating task analysis and model selection for various music tasks. ByteComposer~\cite{liang2024bytecomposer} simulated human composition workflows---concept analysis, draft creation, self-evaluation and revision---aligning AI composition with human creative logic to improve interactivity and interpretability. ComposerX~\cite{composex2024}, built on GPT-4, was the first multi-agent symbolic music composition system, improving quality through role-based collaboration. 

\tmp{However, these systems primarily function as tool integrators and are often constrained by the performance of underlying models, leaving significant room for improvement. 
Music agents hold the potential to transcend these limitations by infusing the robust generative capabilities of platforms like Suno with the advanced planning and reasoning faculties of LLMs. To achieve professional-level autonomous composition, future agents must address the critical hurdle of long-term planning. Unlike linear generation, music demands rigid structural coherence; thus, future agents could adopt hierarchical planning mechanisms, such as recursive LLMs or multi-agent collaboration, to draft global blueprints before micro-generation. Furthermore, incorporating reinforcement learning from human feedback (RLHF) and agentic closed-loop feedback is essential, enabling agents to `listen' to and critique their own generated results for iterative refinement. Crucially, by orchestrating diverse downstream tools, agents could facilitate the synergistic co-creation of symbolic and audio music, combining the structural editability of symbolic representations with the high fidelity of audio synthesis. Finally, expanding beyond composition, agents could integrate music-driven visual generation  (Sec.~\ref{sec:m2v}) and visual-to-music generation (Sec.~\ref{subsec:video-guided-music} ) models to serve as multimodal producers. By integrating visual media, agents could orchestrate complex applications like MV sequence generation or AR concerts, providing intelligent support for the entire creative lifecycle.
}

\section{Datasets and Evaluation}
\label{sec:dataset-evaluation}
The development of cross-modal interaction between music and multimodal data relies on the emergence of diverse datasets involving music and other modalities. Researchers have introduced new evaluation metrics to provide more comprehensive performance validation and support for this field.

\subsection{Datasets}
We have compiled a list of existing multimodal music datasets in Table~\ref{tab:multimodal_datasets}, together with essential descriptions and applicable tasks, providing researchers with a concise overview of available datasets in their areas of interest. Music-text datasets are relatively abundant and form the basis for advances in text-to-music generation and music understanding. In contrast, music-image paired datasets are limited, likely due to the absence of temporal structure in images, which complicates modality alignment. Music-video datasets are an emerging area, yet despite the abundance of music-containing videos on the internet, collecting large-scale, high-quality paired datasets remains challenging. On one hand, many video audio tracks contain noise, such as ambient sounds or speech, in addition to music. On the other hand, due to the ambiguity and diversity of music-video pairing patterns, user-produced videos often feature music that is not well-aligned; more commonly, while semantically harmonious, the music and video lack precise temporal synchronization. Automated filtering methods are underdeveloped, while manual filtering is both expensive and subjective, making large-scale data acquisition difficult.

Early music-video datasets primarily focused on MV collections~\cite{CBVMR2018, mckee2023language, V-MusProd2023}. Recently, driven by the growing demand for general video music generation, newer datasets~\cite{vidmuse2024, zuo2025gvmgen, vidmusician2024} have drawn from a broader range of video scenarios.
Dance videos, a subclass of music-video datasets, are often studied independently. AIST++ is currently the most widely used dataset for music generation from dance. However, its constrained scene diversity and repetitive choreography and songs limit the performance of  dance-to-music generation. The development of more complex and large-scale in-the-wild dance datasets is now a trend in this area.

\subsection{Evaluation}
Table~\ref{tab:metrics} summarizes the objective evaluation metrics currently used in music-related cross-modal tasks, along with brief
{
\setlength{\footskip}{40pt}
\onecolumn
    \small
\setlength{\tabcolsep}{2pt} 

\begin{longtable}{>{\raggedright}p{0.14\textwidth}|>{\raggedright}p{0.15\textwidth}>{\raggedright}p{0.09\textwidth}lp{0.06\textwidth}>{\raggedright}p{0.24\textwidth}>{\raggedright\arraybackslash}p{0.19\textwidth}}
    \caption{Overview of multimodal music datasets.} 
    \label{tab:multimodal_datasets} \\
    \hline
    \tmp{\textbf{Type}} & \textbf{Dataset} & \textbf{Modality} & \textbf{Size} & \tmp{\textbf{Open}} & \textbf{Note} & \textbf{Applications} \\
    \hline
    \endfirsthead
    \caption[]{(continued) Overview of multimodal music datasets.} \\
    \hline
    \tmp{\textbf{Type}} & \textbf{Dataset} & \textbf{Modality} & \textbf{Size} & \tmp{\textbf{Open}} & \textbf{Note} & \textbf{Applications} \\
    \hline
    \endhead
    \hline
    \multicolumn{7}{r}{\small Continued on next page} \\
    \endfoot
    \hline
    \endlastfoot

    \multirow{50}{=}{\tmp{Music-driven cross- modal interaction} }
    & Jamendo Lyrics~\cite{stoller2019end} & audio, text & 80 & \href{https://github.com/f90/jamendolyrics/releases/tag/original}{Yes} &  80 songs with different genres and languages along with lyrics time-aligned on a word-by-word level (with start and end times) to the music & Automatic lyrics transcription (ALT) \& alignment\\ \cline{2-7}
    & MTG-Jamendo~\cite{bogdanov2019mtg} & audio, text & 55k  & \href{http://huggingface.co/datasets/rkstgr/mtg-jamendo}{Yes} & Over 55,000 full audio tracks with 195 tags covering genre, instrument, and mood/theme. & Music auto-tagging \\ \cline{2-7}
    & lyrics-melody dataset~\cite{yu2021conditional} & audio, text &  12,197 & \href{https://github.com/yy1lab/Lyrics-Conditioned-Neural-Melody-Generation}{Yes} & Paired lyrics and melody alignment, from different music sources where the alignment relationship between syllables and music attributes is extracted. & Lyrics generation\\ \cline{2-7}
    & WASABI Song Corpus~\cite{buffa:hal-03282619} & audio, text & 1.73M  & \href{https://github.com/micbuffa/WasabiDataset}{Yes} & A large corpus of songs enriched with metadata extracted from music databases on the Web & Lyrics interpretation \\ \cline{2-7}
    & SongCompose-PT~\cite{ding2024songcomposer}  & audio, text & 283k  & \href{https://huggingface.co/datasets/Mar2Ding/songcompose_data}{Yes} & Large-scale song pretraining dataset that includes lyrics, melodies, and paired lyrics-melodies in either Chinese or English & Lyrics generation\\ \cline{2-7}
    & Chinese-Lyric-Corpus~\cite{gaussic_chinese_lyric_corpus} & text & 50k  & \href{https://github.com/gaussic/Chinese-Lyric-Corpus}{Yes} & Chinese lyrics corpus crawled from NetEase Cloud Music & Lyrics generation\\ \cline{2-7}
    & LP-MusicCaps~\cite{doh2023lp} & audio, text & 22M & \href{https://huggingface.co/datasets/seungheondoh/LP-MusicCaps-MSD}{Yes} & LLM based pseudo music captioning dataset & Music caption \\ \cline{2-7}
    & MusicInstruct~\cite{deng2024musilingo} & audio, text & 5.5k & \href{https://huggingface.co/datasets/m-a-p/Music-Instruct}{Yes} & Music-instruction pairs containing different styles and types & Music understanding\\ \cline{2-7}
    & OpenMU-Bench~\cite{zhao2024openmu} & audio, text & 1M & \href{https://huggingface.co/datasets/Sony/OpenMU-Bench}{Yes} & Extended from existing datasets with rich annotations by GPT-3.5 & Music understanding\\ \cline{2-7}
    & MusicQA~\cite{liu2024music} & audio, text & 12.5k & \href{https://huggingface.co/datasets/mu-llama/MusicQA}{Yes} & Question-answer pairs built on MusicCaps and MagnaTagATune~\cite{law2009evaluation} & Music understanding \\ \cline{2-7}
    & GrooveNet~\cite{GrooveNet2017} & audio, motion & 0.4h & \href{https://github.com/mlab-upenn/GrooveNet}{Yes} & Music-dance paired data collected through Mocap with single dance genre &  Music-to-dance generation \\ \cline{2-7}
    & Dance w/Melody~\cite{TangJM18} & audio, motion & 1.6h & \href{https://github.com/Music-to-dance-motion-synthesis/dataset}{Yes} & Music-dance paired data collected through Mocap in 4 dance genres &  Music-to-dance generation \\ \cline{2-7}
    & Music2Dance~\cite{zhuang2020music2dancedancenetmusicdrivendance} & audio, motion & 1h & \href{https://github.com/zhuang-h/Music2Dance}{Yes} & Music-dance paired data collected through Mocap with hand motion, 2 dance genres &  Music-to-dance generation \\ \cline{2-7}
    & EA-MUD~\cite{SunWCKGL21} & audio, motion & 0.4h & \href{https://github.com/computer-animation-perception-group/DeepDance}{Yes} & Music-dance paired data, 4 dance genres &  Music-to-dance generation \\ \cline{2-7}
    & FineDance~\cite{Li2022FineDanceAF} & audio, motion & 346/14.6h & \href{https://github.com/li-ronghui/FineDance}{Yes} & Music-dance paired data in SMPL-X format, with detailed hand motions, 22 dance genres &  Music-to-dance generation \\ \cline{2-7}
    & PhantomDance~\cite{Li2021DanceFormer} & audio, motion & 260/9.6h & \href{https://github.com/libuyu/PhantomDanceDataset?tab=readme-ov-file#dataset-download}{Yes} & Music-dance paired data in SMPL format, 13 dance genres &  Music-to-dance generation \\ \cline{2-7}
    & POPDG~\cite{Luo_2024_CVPR} & audio, motion & 307/3.5h & \href{https://github.com/Luke-Luo1/POPDG}{Yes} & Music-dance paired data for popular dance in SMPL format, 19 dance genres &  Music-to-dance generation \\ \cline{2-7}
    & AIST-M~\cite{double1} & audio, motion & 1h & \href{https://github.com/JJessicaYao/AIST-M-Dataset}{Yes} & Music-dance paired data of  interactive couple-dances in SMPL format from AIST, 10 dance genres & Duet dance generation \\ \cline{2-7}
    & DD100~\cite{double2} & audio, motion & 2h & \href{https://github.com/lisiyao21/Duolando/tree/main?tab=readme-ov-file}{Yes} & Music-dance paired data of interactive couple-dances in SMPL-X format, 10 dance genres & Duet dance generation \\ \cline{2-7}
    & InterDance~\cite{double4} & audio, motion & 4h & To be released & Music-dance paired data of  interactive couple-dances in SMPL-X format, 15 dance genres & Duet dance generation \\ \cline{2-7}
    & \tmp{MDD~\cite{gupta2025mdd}} & \tmp{audio, motion, text} & \tmp{10.3h} & \tmp{To be released}& \tmp{Music-synchronized 3D couples motion with over 10K text annotations }& \tmp{Text/Music-driven duet dance generation} \\ \cline{2-7}
    & AIOZ-GDANCE~\cite{group1} & audio, motion & 16.7h & \href{https://huggingface.co/datasets/aiozai/AIOZ-GDANCE}{Yes} & Music-dance paired data of group dances in SMPL format, 19 dance genres & Group dance generation \\ \cline{2-7}
    & MDC~\cite{group3} & audio, motion & 73/1h & No & Music-dance paired data of group dances in fbx format & Group dance generation \\ \cline{2-7}
    & SingingHead~\cite{SingingHead} & audio, video, motion & 448, 27h & \href{https://huggingface.co/datasets/Human-X/SingingHead}{Yes} & 2D portrait videos of 76 singers with singing audio, background music and 3D facial motions & Singing head animation \\ \cline{2-7}
    & Song2Face~\cite{Song2Face} & audio, motion & 2h & No & Vocal audio with 3D facial motions (blendshapes) of 7 singers & Singing head animation \\ \cline{2-7}
    & SingingFace~\cite{MusicFace} & audio, motion & 600/40h & No & Singing audio, background music with 3D facial motions and eye state of 7 singers & Singing head animation \\ \cline{2-7}
    & SHV~\cite{Li2024SINGER} & audio, video & 200/20h & To be released & In-the-wild singing video with background music and background image & Singing head animation \\ \cline{2-7}
    & Conductor-Motion100~\cite{liuConducting} & audio, motion & 100h & \href{https://www.scidb.cn/en/detail?dataSetId=305149b2c6654fc7be229f9fa912b8b4}{Yes} & Orchestral conductor motions and aligned music mel spectrogram & Conducting motion generation \\ \cline{2-7}
    & URMP~\cite{li2018creating}& audio, video, MIDI & 44 & \href{https://labsites.rochester.edu/air/projects/URMP.html}{Yes} & Classical chamber music with multi-instrument annotations, frame/note-level transcriptions & Audio-visual synchronization, instrument performance motion generation \\ \cline{2-7}
    & Violin\&Piano Performance Dataset~\cite{Sarasa2017DatasetsFT} & audio, motion & 215 & \href{https://gitlab.doc.gold.ac.uk/expressive-musical-gestures/dataset}{Yes} & Violin and piano audio with performance motion & Instrument performance motion generation \\ \cline{2-7}
    & AudioSet-Drums~\cite{7952261,Cherian2020Sound2SightGV} & audio, video & 8k & \href{https://research.google.com/audioset/ontology/drum_1.html}{Yes} & Drum performance video with visible drummer and clear drum beat from AudioSet & Instrument performance motion generation \\ \cline{2-7}
    & \tmp{MMIP~\cite{kyriakou2025multi}} & \tmp{audio, multi-angle video, motion, MIDI} &\tmp{3.5h} & \tmp{\href{https://mmip.cs.ucy.ac.cy/}{\href{https://github.com/f90/jamendolyrics/releases/tag/original}{Yes}}} & \tmp{Synchronized multi-modal recordings of professional musicians performing on guitar, digital piano, and drums} & \tmp{Instrument performance motion generation} \\ \cline{2-7}
    & Music-AVQA~\cite{li2022learning}, Music-AVQA-R~\cite{ma2024look} & audio, text, video & 9,288 & \href{https://drive.google.com/drive/folders/1WAryZZE0srLIZG8VHl22uZ3tpbGHtsrQ?usp=sharing}{Yes} &  Question-answer pairs about music performance & Audio-visual question answering,  music understanding \\ \hline

    \multirow{8}{=}{\tmp{Music-oriented cross-modal interaction} }
    & MusicCaps~\cite{MusicLM2023} & audio, text & 5.5k & \href{https://huggingface.co/datasets/google/MusicCaps}{Yes} & 10s music examples labeled with aspect lists and captions written by musicians & Text-to-music generation \\ \cline{2-7}
    & MusicBench~\cite{melechovsky2023mustango} & audio, text & 52.8k & \href{https://huggingface.co/datasets/amaai-lab/MusicBench}{Yes} & Extended from MusicCaps & Controllable text-to-music generation \\ \cline{2-7}
    & MUCaps~\cite{liu2024mumu} & audio, text & 18.5k & \href{https://huggingface.co/datasets/M2UGen/MUCaps}{Yes} & 10s music with synthesized language descriptions & Text-to-music generation\\ \cline{2-7}
    & MUSICGEN~\cite{copet2024simple} & audio, text & 400k & No & Instrument-only music tracks & Text-to-music generation \\ \cline{2-7}
    & Stable Audio Open Dataset~\cite{evans2024stable} & audio, text & 486k & \href{https://huggingface.co/stabilityai/stable-audio-open-1.0}{Yes} & CC recordings from Freesound and the Free Music Archive (FMA) & Text-to-music generation\\ \cline{2-7}
    & XMIDI~\cite{tian2025xmusic} & text, MIDI & 108,023 & \href{https://github.com/xmusic-project/XMIDI_Dataset}{Yes} & Largest known symbolic music dataset with precise emotion and genre labels &  Text-to-symbolic music generation \\ \cline{2-7}
    & MUImage~\cite{liu2024mumu} & audio, text, image & 14,520 & \href{https://huggingface.co/datasets/M2UGen/MUImage}{Yes} &  Music samples and corresponding images from the Balanced-AudioSet & Image-to-music generation \\ \cline{2-7}
    & IMEMNet~\cite{zhao2020emotion} & Image, audio & 144k & \href{https://github.com/linkAmy/IMEMNet}{Yes} & Image-music pairs with continuous valence-arousal emotion annotations & Image-driven music retrieval \\ \cline{2-7}
    & EIMG Dataset~\cite{wang2023continuous} & Image, MIDI & 3k & \href{https://github.com/zBaymax/EIMG}{Yes} & Images with VA labels and music clips annotated with continuous valence-arousal values, paired by emotion similarity score in VA space & Image-to-music generation \\ \cline{2-7}
    & EMID~\cite{zou2023emid} & Image, audio & 10,738 & \href{https://huggingface.co/datasets/ecnu-aigc/EMID}{Yes} & Music clips paired with 3 images in the same emotional category, and rich annotations & Image-to-music generation\\ \cline{2-7}
    & MeLBench~\cite{melfusion2024} & image, text, audio & 11,250 & \href{https://umd0-my.sharepoint.com/personal/sanjoyc_umd_edu/_layouts/15/onedrive.aspx?id=%2Fpersonal%2Fsanjoyc%5Fumd%5Fedu%2FDocuments%2FMeLFusion%20datasets&ga=1}{Yes} & Manually annotated triplets of image, text and music, including 15 predefined genres, various image types and detailed text descriptions & Image and text-to-music generation \\ \cline{2-7}
    & Extended MusicCaps~\cite{melfusion2024} & image, text, audio & 7,684 & \href{https://umd0-my.sharepoint.com/personal/sanjoyc_umd_edu/_layouts/15/onedrive.aspx?id=%2Fpersonal%2Fsanjoyc%5Fumd%5Fedu%2FDocuments%2FMeLFusion%20datasets&ga=1}{Yes} & Extended from MusicCaps dataset by adding corresponding images to each text-music pair & Image and text-to-music generation \\ \cline{2-7}
    & Emotion Painting Music Dataset~\cite{hisariya2024bridging} & image, audio & 1,200 & No &Paired paintings and music clips categorized into five emotions (Happy, Angry, Sad, Fun, Neutral) & Image-to-music generation\\ \cline{2-7}
    & Chinese Paintings and Guzheng Audio Dataset (CPGAD)~\cite{zhao2024cptgz} & image, audio & 22,103 & No & Paired dataset of Chinese paintings and guzheng music with semantic annotations & Image-to-music generation \\ \cline{2-7}
    & YouTube8M-MusicTextClips~\cite{mckee2023language} & audio, text, video & 4k & \href{https://zenodo.org/records/8040754}{Yes} & 10s music videos with synthesized language descriptions & Video-to-music recommendation \\ \cline{2-7}
    & HIMV-200k~\cite{CBVMR2018} & audio, video & 200k & \href{https://github.com/csehong/VM-NET}{Yes} & Music videos derived from the YouTube-8M dataset & Video-to-music recommendation \\ \cline{2-7}
    & TT-150k~\cite{corssmodal2021}& audio, text, video & 150k & No & Micro-videos based on 3k music & Video background music recommendation \\ \cline{2-7}
    & SymMV~\cite{V-MusProd2023} & audio, video, MIDI & 1,140 & \href{https://github.com/zhuole1025/SymMV/tree/main/dataset}{Yes} &Music videos with rich musical annotations & Video-to-music generation \\ \cline{2-7}
    & BGM909~\cite{li2024diff} & audio, text, video, MIDI & 909 & \href{https://github.com/sizhelee/Diff-BGM}{Yes} & Music videos with rich annotations & Video-to-music generation \\ \cline{2-7}
    & MUVideo~\cite{liu2024mumu} & audio, text, video & 14,504 & \href{https://huggingface.co/datasets/M2UGen/MUVideo}{Yes} &  Music samples and corresponding videos from the Balanced-AudioSet & Video-to-music generation \\ \cline{2-7}
    & MuVi-Sync~\cite{kang2024video2music} & audio, video & 748 & \href{https://zenodo.org/records/10057093}{Yes} & Music videos with rich features (chords, tonality, scene offsets) &Video-to-music generation \\ \cline{2-7}
    & V2M~\cite{vidmuse2024}& audio, video & 360k & \href{https://huggingface.co/datasets/HKUSTAudio/VidMuse-V2M-Dataset}{Yes} & Diverse video-music pairs & Video-to-music generation \\ \cline{2-7}
    & GVMGen~\cite{zuo2025gvmgen}& audio, video & 147 h & \href{https://github.com/chouliuzuo/GVMGen}{Yes} & Diverse video-music pairs including Chinese traditional music & Video-to-music generation\\ \cline{2-7}
    & DVMSet~\cite{vidmusician2024} & audio, video & 3,839& To be released & Diverse video-music pairs &Video-to-music generation \\ \cline{2-7}
    & MusicPro-7k~\cite{xie2025filmcomposer} & audio, text, video & 7,418 & \href{https://huggingface.co/datasets/apple-jun/MusicPro-7k}{Yes} & Film music dataset with film clips, music, visual and music descriptions, melodies, and rhythm spots per sample & Film-to-music generation \\ \cline{2-7}
    & FilmScoreDB~\cite{qi2024harmonizing} & video, music & 32,520 & To be released & 90.35 hours of film clips and  corresponding  original scores by renowned film composers & Film-to-music generation \\ \cline{2-7}
    & HarmonySet~\cite{zhou2025harmonyset} & audio, video & 48,328 & \href{https://huggingface.co/datasets/Zzitang/HarmonySet}{Yes} & Diverse video-music pairs & Video-to-music generation \\ \cline{2-7}
    & LORIS~\cite{yu2023long} & audio, video & 12,446 & \href{https://huggingface.co/datasets/awojustin/LORIS}{Yes} &25 s  paired videos, including 8,585 figure skating and 1,950 floor exercise clips  & Sports-to-music generation \\ \cline{2-7}
    & TikTok~\cite{V2M-GAN2022} & audio, video & 445 & \href{https://github.com/L-YeZhu/D2M-GAN}{Yes} & In-the-wild dance videos with complex backgrounds & Dance-to-music generation \\ \cline{2-7}
    & Suno Music Generation Dataset~\cite{suno_music_generation_dataset} & audio, text, video, image & 660k  & \href{https://huggingface.co/posts/nyuuzyou/391290114296515}{Yes} & Metadata for 659,788 songs generated by artificial intelligence on the suno.com platform &Any-to-music generation\\
    \hline

    \multirow{25}{=}{\tmp{Bidirectional cross- modal interaction} }
    & AIST++~\cite{AIST} & audio, video & 1,408 & \href{https://google.github.io/aistplusplus_dataset/}{Yes} & Multi-view dance dataset, 10 dance genres & Music-dance mutual generation \\ \cline{2-7}
    & AIST-M2B~\cite{Jiang2024M2BeatsWM} & audio, motion & 1408 & \href{https://github.com/mRobotit/M2Beats}{Yes} & Algorithm for extracting motion rhythm labels from music-dance paired data in the AIST++ dataset & Motion rhythm analysis, dance-music alignment \\ \cline{2-7}
    & Dance50~\cite{Yu2022LearningMR} & audio, video & 14k, 47h & No & K-pop dance clips collected from website with 2D skeleton keypoints extracted by OpenPose & Motion rhythm analysis, Dance-music alignment \\ \cline{2-7}
    & AnyInstruct~\cite{zhan2024anygpt} & audio, text, image & 108k & \href{https://huggingface.co/datasets/OpenMOSS-Team/AnyInstruct}{Yes} & Multimodal interleaved instruction-following data, constructed by a generative model & Any-to-any generation \\ \cline{2-7}
    & MMtrail~\cite{chi2024mmtrail} & audio, text, video & 20M & \href{https://huggingface.co/datasets/litwell/MMTrail-20M}{Yes} & 27k+ hours of trailer clips, multi-modal captions &  Multimodal understanding and generation \\ \cline{2-7}
    & EmoMV~\cite{thao2023emomv} & audio, video & 5,986 & \href{https://doi.org/10.5281/zenodo.7011072}{Yes} & Humanly and automatically annotated labels for emotional correspondence & Music-video correspondence learning \\ \cline{2-7}
    & MusicPile-sft~\cite{chatmusician2024} & text, ABC notation & 1,139,473 & \href{https://huggingface.co/datasets/m-a-p/MusicPile-sft}{Yes} & Samples with a ratio of music verbal to  score of 2:1, created using music from various regions &  Music understanding and generation \\ \cline{2-7}
    & MUEdit~\cite{liu2024mumu}  & audio, text & 5k & \href{https://huggingface.co/datasets/M2UGen/MUEdit}{Yes} & Instruction-response pairs for music and text & Instruction-based music editing\\
    \hline

    \multirow{8}{=}{\tmp{Music-driven \& Music-oriented interaction}}
    & Youtube-Music-1M~\cite{zhan2024anygpt} & audio, text & 1M & \href{https://huggingface.co/datasets/OpenMOSS-Team/AnyInstruct}{Yes} & Music-text pairs from Youtube &  Music captioning, text-to-music generation\\ \cline{2-7}
    & WikiMusicText (WikiMT)~\cite{wu2023clamp} & ABC, text & 1010 & \href{https://huggingface.co/datasets/sander-wood/wikimusictext}{Yes} &  Sourced from Wikifonia.org, lead sheets with title, artist, genre, and description. & Text-to-music generation, music captioning \\ \cline{2-7}
    & Song Describer Dataset (SDD)~\cite{manco2023song} & audio, text & 706 & \href{https://github.com/mulab-mir/song-describer-dataset}{Yes} & High-quality benchmark for evaluating music-and-language models & Music captioning, text-to-music generation, text-to-music retrieval \\ \hline

\end{longtable}

\setlength{\tabcolsep}{3pt}
\begin{longtable}{>{\raggedright\arraybackslash}p{0.15\textwidth}|>{\raggedright\arraybackslash}p{0.2\textwidth}>{\raggedright\arraybackslash}p{0.42\textwidth}>{\raggedright\arraybackslash}p{0.18\textwidth}}
    \caption{Overview of objective evaluation metrics for music-related cross-modal tasks.}
    \label{tab:metrics}\\
    \hline
    \textbf{Type} & \textbf{Metric} & \textbf{Description} & \textbf{Applications} \\ 
    \hline
    \endfirsthead
    \caption[]{(continued) Overview of objective evaluation metrics for music-related cross-modal tasks. } \\
    \hline
    \textbf{Type} & \textbf{Metric} & \textbf{Description} & \textbf{Applications} \\ 
    \hline
    \endhead
    \hline
    \multicolumn{4}{r}{\small Continued on next page} \\
    \endfoot
    \hline
    \endlastfoot
    \multirow{30}{*}{\parbox{0.15\textwidth}{Music-driven cross-modal interaction}}
     & Accuracy 
    & Ratio of correctly classified instances to total number of instances 
    &Multimodal representation learning \\
    \cline{2-4}
    & Recall 
    & Ratio of all relevant items successfully retrieved in the top-$K$ results 
    &Multimodal representation learning \\
    \cline{2-4}
    & Mean average precision (mAP)
    & Mean of average precision over all queries, measuring both relevance and ranking quality 
    & Multimodal representation learning \\
    \cline{2-4}

     & BLEU~\cite{papineni2002bleu}
    & $n$-gram overlap between generated text and reference answers 
    & Music understanding \\
    \cline{2-4}
    & Meteor~\cite{banerjee2005meteor} 
    & Text similarity  considering precision, recall, and alignment based on exact, stem, synonym, and paraphrase matches
    & Music understanding  \\
    \cline{2-4}
    & Rouge-L~\cite{lin2004rouge}
    & Text similarity by  longest common subsequence (LCS), capturing semantic continuity and word order
    & Music understanding \\
    \cline{2-4}
    & BERT-score~\cite{zhangbertscore}
    & Semantic similarity using BERT embeddings of generated and reference text 
    & Music understanding \\
    \cline{2-4}
    & Word error rate (WER)
    & Edit distance (Levenshtein distance) between the hypothesis (predicted transcription) and the reference (ground-truth transcript) 
    & Lyrics transcription \\
     \cline{2-4}
     & Beat align score (BAS)~\cite{AIST}
    & Motion-music correlation using the average distance between each kinematic beat and its nearest music beat 
    & Music-to-motion generation \\
    \cline{2-4}
    & Diversity (Dist)~\cite{AIST}
    & Generation diversity using the average Euclidean distance in  feature space across generated motions
    & Music-to-motion generation \\
    \cline{2-4}
    & Landmark distance (LMD)~\cite{Chen2018LipMG}
    & Evaluation of lip synchronization using  Euclidean distance between mouth landmarks between ground-truth and generated motions
    & Music-to-motion generation \\
    \cline{2-4}
    & Frechét inception distance (FID)~\cite{AIST}
    & Motion quality using FID based on geometric and kinetic features between ground-truth(GT) and generated motions
    & Music-to-motion generation \\
    \cline{2-4}
    & Standard structural similarity (SSIM)~\cite{10.1109/TIP.2003.819861}
    & Quality of  generated video frames compared to ground-truth using  differences between the properties of the pixels
    & Music-to-video generation \\
    \cline{2-4}
    & Signal to noise ratio (PSNR)
    & Quality of  generated video frames compared to ground-truth  using  absolute error between  pixels.
    & Music-to-video generation \\
    \hline
    
    \multirow{20}{*}{\parbox{0.15\textwidth}{Music-oriented cross-modal interaction }}
   & Frechét audio distance (FAD) 
    & Similarity of distributions of generated audio and real music based on VGGish~\cite{hershey2017cnn} 
    & Any-to-music generation  \\
    \cline{2-4}
    & Frechét distance (FD) 
    & Similarity between  distributions of generated audio and real music based on PANNS~\cite{kong2020panns} or OpenL3~\cite{cramer2019look} 
    & Any-to-music generation \\
    \cline{2-4}
    &Kullback-Leibler divergence (KL) 
    & Divergence between  class predictions of ground truth and generated music based on PANNS~\cite{kong2020panns} or PaSST~\cite{koutini2021efficient}
    & Any-to-music generation  \\
    \cline{2-4}
     & Inception score (IS) & Diversity and quality of generated music, measured by  entropy of class predictions using a pre-trained music classifier (e.g., VGGish~\cite{hershey2017cnn} or PANNS~\cite{kong2020panns}) & Any-to-music generation \\ 
     \cline{2-4}
    & CLAP~\cite{large2022} score & Similarity of
      CLAP embeddings of music and text & Text-to-music generation \\
     \cline{2-4}
     & ClaMP3~\cite{wu2025clamp} score & A more recent metric similar to CLAP score & Text-to-music generation \\
     \cline{2-4}
     & Phoneme error rate (PER)
     & Phoneme recognition accuracy using FireRedASR~\cite{xu2025fireredasr}
     & Any-to-song generation\\
     \cline{2-4}
     & Vocal range
     & Quantifies the breadth of the vocal range at the song level (pitch estimated with RMVPE~\cite{RMVPE})
     & Any-to-song generation \\
     \cline{2-4}
    & Imagebind score~\cite{girdhar2023imagebind} & Similarity of ImageBind embeddings of music and video & Video-to-music generation \\
    \cline{2-4}
    &Beats coverage score (BCS)
    & Ratio of  generated beats to  ground truth beats
    & Dance-to-music generation\\
    \cline{2-4}
    &Beats hit score (BHS)
    &Ratio of  aligned generated beats to number of  ground truth beats
    & Dance-to-music generation\\
    \cline{2-4}
    &Average sample-wise accuracy (ASA)
    & Proportion of correctly predicted attributes per sample, averaged over the test set
    & Text-to-music generation\\
    \cline{2-4}
     & Tone span (TS) 
    & Number of half-tone steps between the lowest and highest tones in a melody 
    & Any-to-symbolic-music generation\\
    \cline{2-4}
    & Scale consistency (SC) 
    & The fraction of tones in a melody that belong to standard musical scales
    & Any-to-symbolic-music generation\\
    \cline{2-4}
    & Polyphony rate (PR) 
    & The fraction of time steps with multiple simultaneous pitches, measuring polyphonic complexity
    & Any-to-symbolic-music generation\\
    \cline{2-4}
    & Pitch entropy (PE) 
    & Quantification of pitch diversity using information entropy
    & Any-to-symbolic-music generation\\
    \cline{2-4}
    & Groove consistency (GC) 
    & Rhythmic stability = $1-$  average Hamming distance of onset vectors between consecutive measures
    & Any-to-symbolic-music generation\\
    \cline{2-4}

    &Repetition rate
    &how often the repeat sign $:|$ appears in a generated set
    & Any-to-symbolic-music generation\\
    \cline{2-4}
    &Intra similarity
    &Average value of  texture latent similarity matrix, excluding the diagonal
    & Any-to-symbolic-music generation\\
    \cline{2-4}
    &Time signature error rate (TSER)
    &Assesses whether the beat count in each measure corresponds to the time signature in the ABC notation header
    & Any-to-symbolic-music generation\\
    \cline{2-4}
    &Instrument range error rate (IRER)
    &Assesses how often  generated notes exceed the expert-defined instrument range
    & Any-to-symbolic-music generation\\
    \cline{2-4}
    &Score information completeness rate (SICR)
    &Assesses completeness of the ABC notation
    & Any-to-symbolic-music generation\\
    \cline{2-4}
    &Average attribute accuracy (AAA)
    & How often attributes in the generated ABC notation match those defined, averaged over all musical attributes
    & Any-to-symbolic-music generation\\
    \cline{2-4}
    &Hit rate@$K$
    &Proportion of top-$K$ recommended songs that contain keywords or  conceptually similar terms associated with the input
    & Any-to-music retrieval\\
    \cline{2-4}
    &Recall@$K$
    &Proportion of relevant items retrieved within the top-$K$ results
    &Any-to-music retrieval\\
    \cline{2-4}
    &Entropy (H@$K$)
    & Distribution of ground-truth genres using the formula on the top-$K$ results
    &Any-to-music retrieval\\
    \cline{2-4}
    & Emotion matching (EM) 
    & Average inverse Euclidean distance between predicted music VA values and image VA values in the valence-arousal space
    & Image-to-music generation\\
    \cline{2-4}
    & Image music similarity metric (IMSM) 
    & The similarity between image-music pairs via CLIP image-text and CLAP audio-text embeddings
    & Image-to-music generation \\
    \cline{2-4}
    & Mel-cepstral distortion (MCD) 
    & The mean Euclidean distance of mel-cepstral coefficients, measuring  spectral distortion between synthesized and original voices
    & Singing voice synthesis/conversion \\
    \cline{2-4}
    & Root-mean-square error (RMSE) of $F_0$ values
    & The square root of the mean of squared errors between the natural and synthesized $F_0$ (fundamental frequency) values, measuring pitch prediction accuracy
    & Singing voice synthesis \\
    \cline{2-4}
    & Correlation coefficients (Corr.) of $F_0$ values
    & Linear correlation coefficient relating natural and synthesized $F_0$ values
    & Singing voice synthesis \\
    \cline{2-4}
    & Duration root mean square error (Dur RMSE)
    & Root mean square error of phoneme durations
    & Singing voice synthesis \\
    \cline{2-4}
    & Duration correlation (Dur CORR)
    & Correlation coefficient relating predicted phoneme duration and the actual duration
    & Singing voice synthesis \\
    \cline{2-4}
    & Real-time factor (RTF)
    & Ratio of times of transforming embeddings into latent representations/mel-spectrograms to the duration of audio, measuring inference speed
    & Singing voice conversion \\
    \cline{2-4}
    & Character error rate (CER)
    & Accuracy of lyrics in  converted audio, using Whisper
    & Singing voice conversion \\
    \cline{2-4}
    & Speaker similarity (SIM)
    & Cosine distance of speaker embeddings, measuring  similarity of  singer's timbre
    & Singing voice conversion \\
    \cline{2-4}
    & $F_0$ Pearson correlation coefficient (FPC)
    & Correlation between  $F_0$ values of  original and converted audio, reflecting the accuracy of pitch conversion
    & Singing voice conversion \\

    \hline
    Bidirectional cross-modal interaction in music 
    & Chroma similarity 
    & Cosine similarity of chromagrams of  original and edited music
    & Music editing \\
    \hline
    
\end{longtable}
\twocolumn
}
descriptions and their associated tasks. \tmp{
Given the diversity of tasks, we analyze the significance of these metrics in four critical ways. First, for tasks like music retrieval and captioning, standard information retrieval metrics (e.g.\ accuracy, mAP) and NLP metrics (e.g. BLEU, WER) serve as the fundamental baseline for quantifying a model's ability to identify musical attributes or transcribe lyrics with linguistic precision. Second, in the domain of audio and video generation, distribution-based metrics like {FAD} and {IS} quantify the statistical distance between generated and reference feature distributions. They are critical for jointly assessing fidelity and diversity, ensuring that the generated signals align with the perceptual characteristics of real-world data. Third, interaction tasks require dual alignment checks: semantically, metrics like {CLAP score} quantify the semantic correspondence (e.g.\ style, and emotional alignment) between music and text, while temporally, metrics such as {BCS} are indispensable for ensuring visual movements synchronize with musical rhythms. Finally, specific to symbolic generation, domain-specific metrics like {pitch entropy} and {groove consistency} are essential to validate adherence to harmonic rules and structural patterns.}

However, unlike the well-established metrics for assessing musical attributes like quality, measures for evaluating cross-modal alignment between music and other modalities are still in the exploratory stage.
A prevalent approach involves extracting features using contrastive learning-based multimodal pretraining models~\cite{large2022, wu2025clamp, girdhar2023imagebind}, and computing similarity scores across modalities. However, most of these models are trained on general audio data rather than music-specific datasets, resulting in limited representational capacity for music and often overlooking temporal characteristics.
Moreover, due to the artistic nature of music, its alignment with other modalities is inherently more flexible. For example, a single video may be effectively paired with multiple music tracks, each creating a distinct but coherent audiovisual experience. This raises an open challenge: how to design evaluation metrics that balance tolerance for flexible pairings with the need for alignment accuracy.

For subjective evaluation, \tmp{human assessment remains the gold standard. Given that music is an art form driven by human perception, objective metrics often fail to capture aesthetic nuances, emotional resonance, and culturally appropriate stylistic attributes.} Researchers typically define task-specific criteria and conduct user studies. These primarily take two forms: rating-based evaluations, such as the human mean opinion score (MOS), where users assign scores to different aspects (e.g., semantics, rhythm, overall quality) of a sample; and preference-based evaluations, where users choose between options, yielding rates that reflect comparative preferences.

\section{Discussion}
\label{sec:discussion-summary}

Despite progress in research involving music and other modalities, notable challenges remain. One critical issue is the lack of high-quality multimodal music datasets, especially ones with paired music–visual data. For example, although numerous music videos exist online, many of them suffer from poor quality. On the other hand, most existing works on cross-modal interactions of music focus on instrumental music, while a large portion of online music takes the form of songs, leading to issues such as the limited availability of instrumental music and data imbalance between music genres. While source separation models can extract instrumental tracks from songs, their limited performance introduces noise into the audio. 

Additionally, the subjective nature of cross-modal alignment and varying levels of users' creative abilities contribute to inconsistencies in internet-sourced data. Manual annotation is also time-consuming and requires expert knowledge.

Another limitation is that current research typically treats pure instrumental music and vocal music separately; most cross-modal music interaction studies are based on instrumental music. However, directly generating full songs is both more efficient and better aligned with the way music is commonly represented on the Internet, making it more user-friendly in practical applications.

Moreover, existing models often exhibit deficiencies in the depth and breadth of modality fusion when handling multimodal data. Many approaches simply concatenate features from different modalities or force them into a shared latent space without truly guiding the model to comprehend the semantic associations and logical relationships between modalities. 
As a result, there is a lack of high-performance music understanding models and multimodal models tailored for music, comparable to the performance of vision-language models (VLMs) in the visual domain.

Furthermore, music agents also face certain limitations. While they have achieved progress in tasks such as music composition assistance and basic music analysis, their performance in complex music understanding and generation remains weak. 
When processing multimodal inputs, music agents may suffer from information loss, leading to decision-making and outputs that lack coherence. Meanwhile, current music agents lack effective feedback mechanisms to enable self-improvement or alignment with personalized preferences.
Moreover, current music agents are based on fixed, manually designed workflows. They lack deep integration with large language models and efficient integration of diverse music tools, limiting their ability to autonomously plan workflows and handle various music processing tasks.

Despite current limitations, several promising trends are emerging in this field:
\begin{enumerate}
\item \emph{High-quality data construction:} advances in technology will improve multimodal data collection and automated annotation, yielding richer, high-quality music data. Automated annotation benefits from the synergy between music generation and understanding, further augmented by LLMs. 
\item \emph{Adaptation of speech techniques:} music and speech share fundamental signal representations (e.g., spectrograms) and generative backbones (e.g., Transformers). This implies significant potential for transferring acoustic priors from large-scale speech models to music tasks. However, bridging the domain gap requires more than direct application; future research should incorporate hierarchical tokenization to decouple high-level semantics from acoustic nuances, long-context memory mechanisms to accommodate music's extended temporal nature, and relative positional encodings to ensure strict alignment with musical bars and beats.
\item \emph{Hybrid symbolic-audio frameworks:} while end-to-end audio generation dominates in acoustic fidelity, symbolic generation remains indispensable for its explicit structure and editability. Unlike raw audio, symbolic representations offer transparent formats allowing precise refinement. Consequently, the evolving paradigm points towards hybrid models that leverage symbolic priors for structural control and neural synthesizers for high-fidelity rendering. This synergy fosters artist-friendly systems, enabling interactive editing and guiding AI music generation towards professional-grade quality.
\item \emph{Multimodal CoT:} the construction of comprehensive multimodal music CoT datasets holds promise for helping models learn logical correspondences across different modalities.
\\item \emph{LLM-Driven multimodal integration:} researchers are exploring novel model architectures and leveraging LLMs to improve multimodal integration, thereby capitalizing on the consistency between music and other modalities for controllable cross-modal generation and understanding.
\item \emph{Alignment with human preferences:} preference data can also be constructed and utilized through methods such as DPO to align generative models more closely with human preferences.
\end{enumerate}

\section{Conclusions}
This paper has provided a comprehensive review of research into cross-modal interaction between music and multimodal data. Based on an overview of music data formats, it systematically examines the development status of music-driven cross-modal interaction, music-oriented cross-modal interaction, and bidirectional cross-modal music interaction. In addition, we comprehensively summarize  existing multimodal music data and the evaluation metrics used in the cross-modal interactions between music and other modalities.
Furthermore, we discuss existing research limitations and future research directions. This survey aims to serve as a  reference for researchers in the field, fostering further advances in cross-modal interaction between music and multimodal data, and driving innovation in the application of related technologies in the music industry and beyond.

\section*{Declaration}
\subsection*{Declaration of competing interests}
The authors have no competing interests to declare that are relevant to the
content of this article.

\subsection*{Author contributions}
Sifei Li and Mining Tan were primarily responsible for drafting and revising the manuscript. Minyan Luo and Feier Shen wrote Sections 3.1 and  3.3, respectively. Zijiao Yin contributed to  proofreading of the manuscript. Weiming Dong led the project. Fan Tang, Weiming Dong, and Changsheng Xu provided overall guidance and valuable suggestions throughout the work.

\subsection*{Acknowledgements}
This work was supported by the National Natural Science Foundation of China (62572458).


\bibliographystyle{CVMbib}
\bibliography{refs}

\subsection*{Author biographies}
\begin{biography}[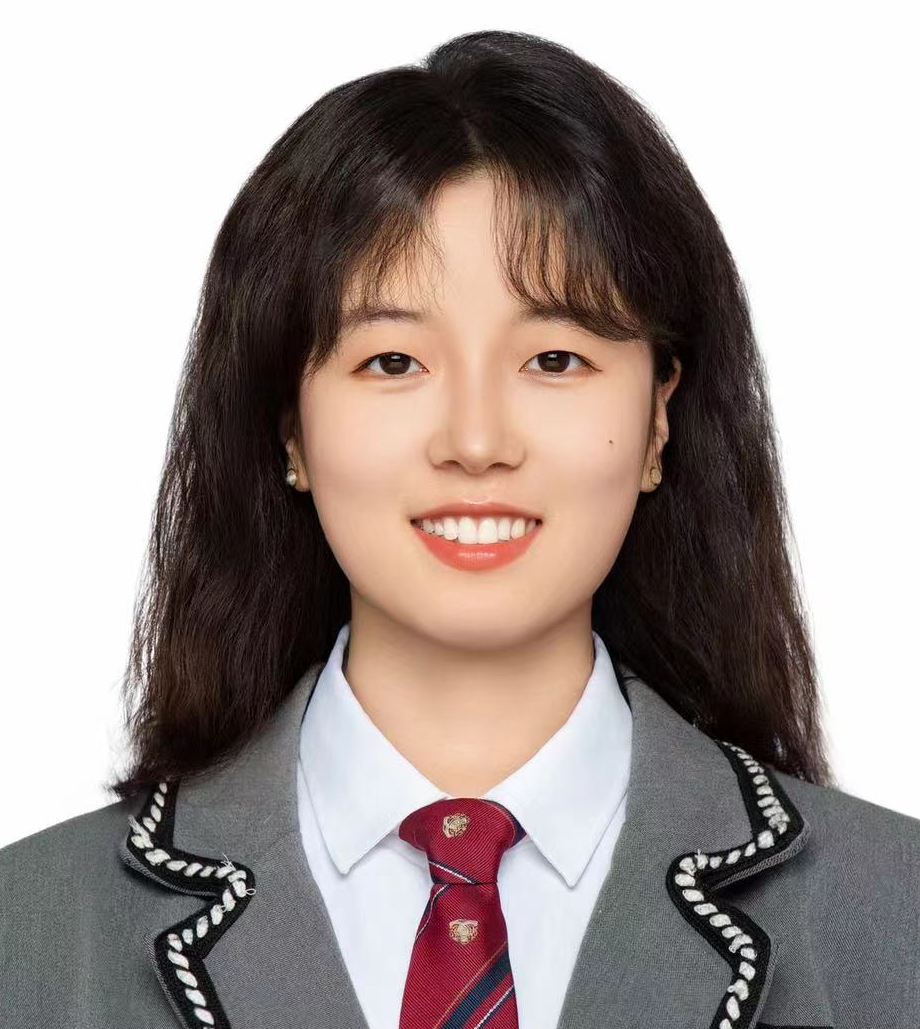]{Sifei Li} received a B.S.  in automation in 2022 from Shandong University, China. She is pursuing a
Ph.D. degree in  the State Key Laboratory of Multimodal Artificial Intelligence Systems (MAIS), Institute of Automation, Chinese Academy of Sciences. Her research interests
include multimodal music generation and computational art.
\end{biography}
\vspace*{1.2em}
\begin{biography}[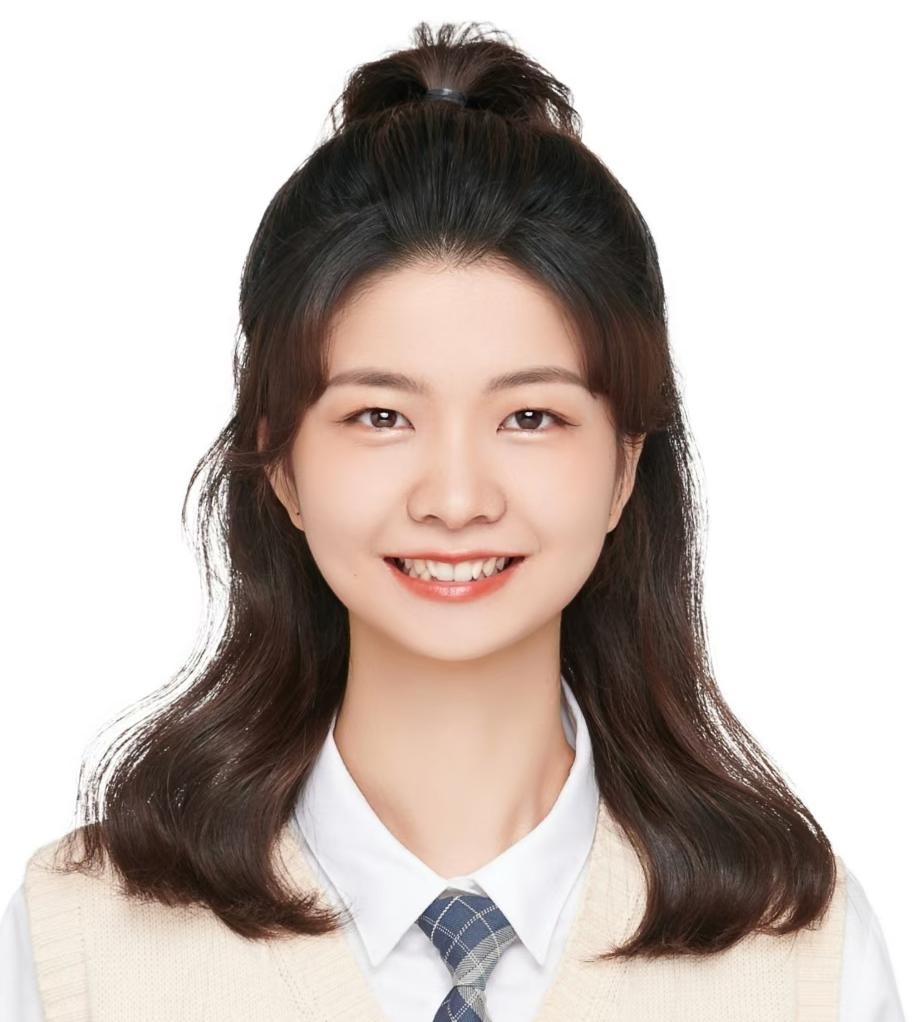]{Mining Tan} received a B.S.  in automation from the University of Science and Technology, Beijing, China, in 2024. She is working towards an M.S. degree in control theory and control engineering in the State Key Laboratory of Multimodal Artificial Intelligence Systems, Institute of Automation, Chinese Academy of Sciences. Her current research interests include deep learning and multimedia computing.
\end{biography}
\vspace*{1.2em}
\begin{biography}[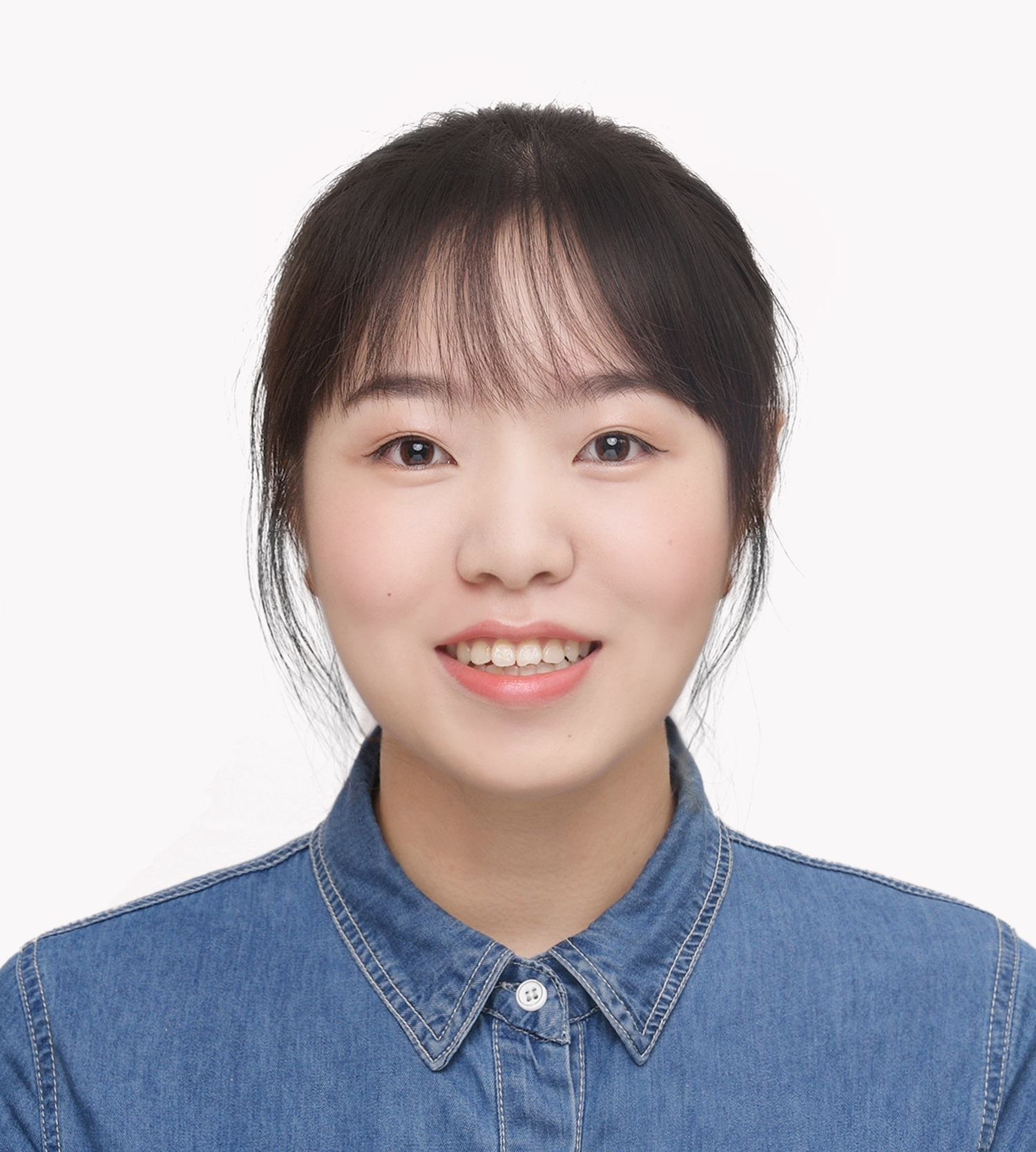]{Feier Shen} received the B.S. degree in Artificial Intelligence in 2025 from the University of Chinese Academy of Sciences, China. She is a Ph.D. student at the State Key Laboratory of Multimodal Artificial Intelligence Systems (MAIS), Institute of Automation, Chinese Academy of Sciences. Her research interests include motion generation and computational art.
\end{biography}
\vspace*{1.2em}
\begin{biography}[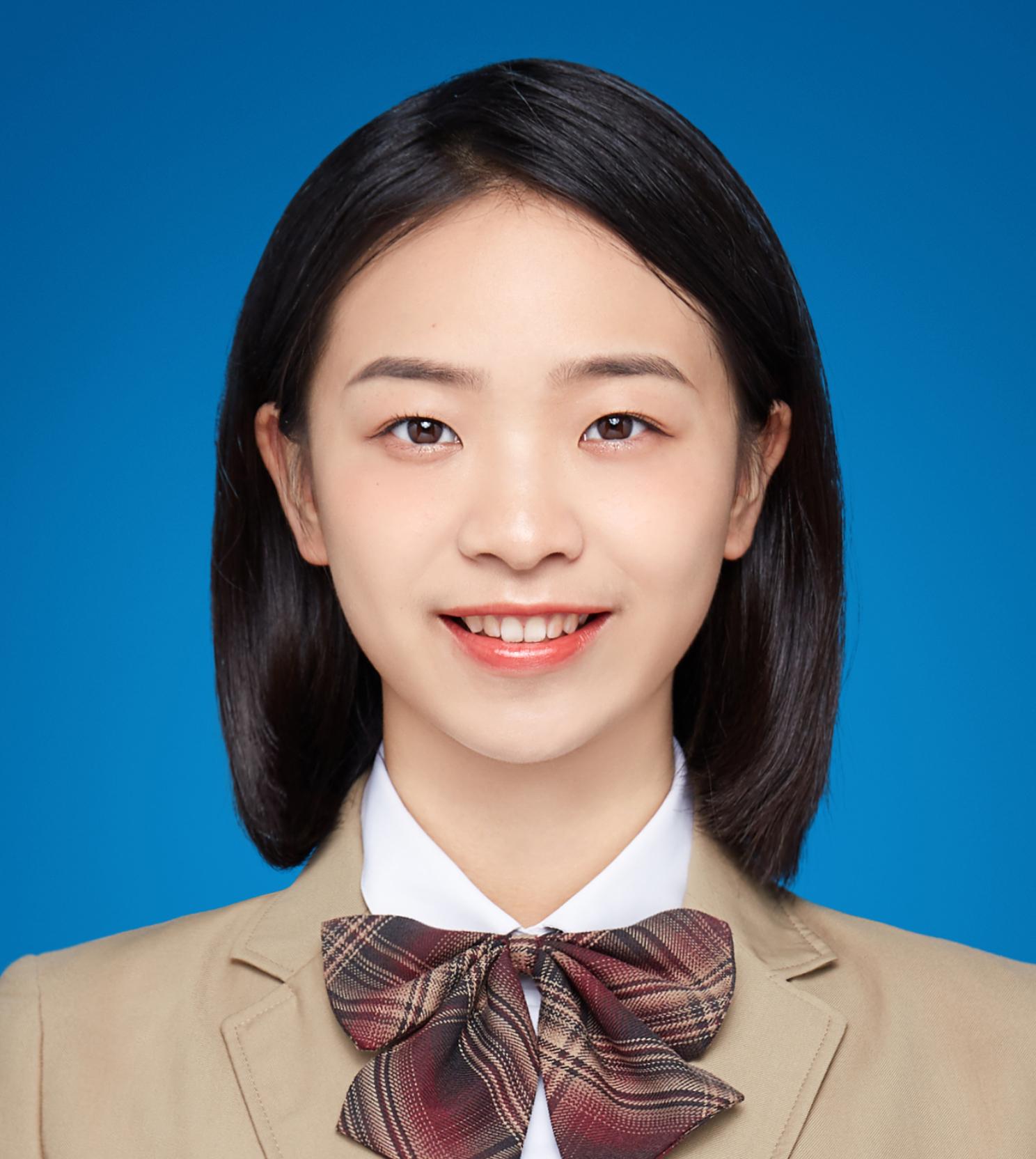]{Minyan Luo} received the B.S. degree in Artificial Intelligence in 2025 from the University of Chinese Academy of Sciences, China. She is a Ph.D. student at the State Key Laboratory of Multimodal Artificial Intelligence Systems (MAIS), Institute of Automation, Chinese Academy of Sciences. Her research interests include computer vision, computer graphics, and computational art.
\end{biography}
\vspace*{1.2em}
\begin{biography}[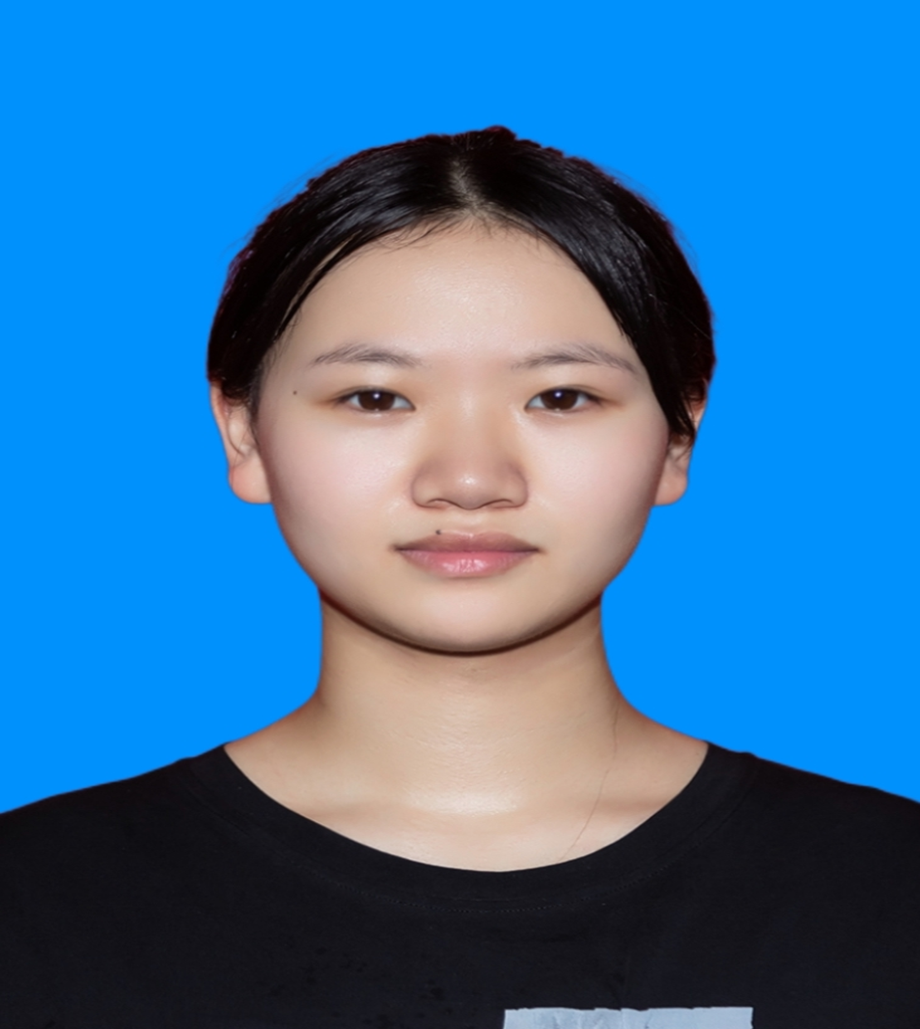]{Zijiao Yin}is an undergraduate student at the University of Chinese Academy of Sciences, majoring in Artificial Intelligence, and will receive her B.Sc. degree in 2025. She will continue her studies as a Ph.D. student at the State Key Laboratory of Multimodal Artificial Intelligence Systems (MAIS), Institute of Automation, Chinese Academy of Sciences. Her research interests include computer vision, computer graphics, and computational art.
\end{biography}
\vspace*{1.2em}
\begin{biography}[./figures/authors/fantang]{Fan Tang} received a B.Sc.  in computer science from North China Electric Power University, Beijing, China, in 2013, and a Ph.D. degree from the Institute of Automation, Chinese Academy of Sciences, Beijing, in 2019. He is an Assistant Professor in the Institute of Computing Technology, Chinese Academy of Sciences. His research interests include computer graphics, computer vision, and machine learning.
\end{biography}
\vspace*{1.2em}
\begin{biography}[./figures/authors/dongweiming]{Weiming Dong} is a Professor in MAIS. He received his B.Sc. and M.Sc. degrees in 2001 and 2004, both from Tsinghua University, China. He received his Ph.D. in Computer Science from the University of Lorraine, France, in 2007. His research interests include artificial intelligence generated content and computational art.
\end{biography}
\vspace*{1.2em}
\begin{biography}[./figures/authors/changshengxu]{Changsheng Xu} is a Professor in MAIS. He received the Best Associate Editor Award of ACM Transactions on Multimedia Computing, Communications, and Applications in 2012 and the Best Editorial Member Award of ACM/Springer Multimedia Systems Journal in 2008. He has served as in various editorial roles for over 20 prestigious IEEE and ACM  multimedia journals, conferences, and workshops. He is an IEEE Fellow, IAPR Fellow and an ACM Distinguished Scientist.
\end{biography}

\end{document}